\documentclass[12pt,oneside,a4paper]{article}
\usepackage{hyperref}
\hypersetup{colorlinks,bookmarksopen,bookmarksnumbered,citecolor=blue,
linkcolor=black,pdfstartview=FitH,urlcolor=blue}

\usepackage{graphicx}
\usepackage{amssymb}
\usepackage{cite}
\usepackage{bm}
\usepackage{indentfirst}
\usepackage{amsmath}
\usepackage{hhline}
\usepackage{multirow}
\usepackage{color}

\allowdisplaybreaks
\numberwithin{equation}{section}

\renewcommand{\bar}{\overline}

\definecolor{dred}{rgb}{0.7,0.15,0.09}
\definecolor{dblue}{rgb}{0,0.12,0.64}
\definecolor{dgreen}{rgb}{0.2,0.51,0.19}

\newcommand{\1}{\mbox{1}\hspace{-0.25em}\mbox{l}}

\begin{document}

\begin{titlepage}

\begin{flushright}
KANAZAWA-22-04
\end{flushright}

\begin{center}

\vspace{1cm}
{\large\textbf{
Impact of first-order phase transitions on dark matter production in the scotogenic model
}
 }
\vspace{1cm}

\renewcommand{\thefootnote}{\fnsymbol{footnote}}
Hiroto Shibuya$^{1}$\footnote[1]{h\_shibuya@hep.s.kanazawa-u.ac.jp}
and 
Takashi Toma$^{1,2}$\footnote[2]{toma@staff.kanazawa-u.ac.jp}
\vspace{5mm}

\textit{
 $^1${Institute for Theoretical Physics, Kanazawa University, Kanazawa 920-1192, Japan}\\
 $^2${\mbox{Institute of Liberal Arts and Science, Kanazawa University, Kanazawa 920-1192, Japan}}
}

\vspace{8mm}

\abstract{
\noindent
In this work, we investigate the effects of first-order phase transitions on the singlet fermionic dark matter in the scotogenic model. 
It is known that this dark matter candidate tends to conflict with the relevant constraints such as the neutrino oscillation data and charged lepton flavor violating processes 
if its thermal production mechanism is assumed. 
We find that the dark matter production mechanisms are modified by first-order phase transitions at some specific parameter regions, 
where the phase transitions can be one-step or two-step depending on the parameters. 
If the phase transition is one-step, a sufficiently low nucleation temperature is required to reproduce the observed relic abundance of dark matter. 
If the phase transition is two-step, the dark matter should never be thermalized, otherwise the abundance would remain too much and overclose the universe. 
This is because the nucleation temperature cannot be low as in the one-step case. 
Therefore we require another way of dark matter production, the freeze-in mechanism for the two-step case. 
We show that the freeze-in mechanism is modified by the temporary vacuum expectation value of the inert scalar field. 
In both cases, the first-order phase transitions could produce observable gravitational wave spectra. 
In particular for the one-step phase transition, the generated gravitational waves with sizable energy density are intrinsically correlated with the dark matter production mechanism, and can be detectable by future space-based interferometers.
}

\end{center}
\end{titlepage}

\renewcommand{\thefootnote}{\arabic{footnote}}
\newcommand{\bhline}[1]{\noalign{\hrule height #1}}
\newcommand{\bvline}[1]{\vrule width #1}

\setcounter{footnote}{0}

\setcounter{page}{1}
%%%%%%%%%%%%%%%%%%%%%%%%%%%%%%%%%%%%%%

\section{Introduction}
The freeze-out~\cite{Kolb:1990vq} and freeze-in mechanisms~\cite{Hall:2009bx} are the standard ways of generating the observed dark matter relic abundance. 
In the former case, the dark matter is kept in thermal equilibrium via sufficient interactions with Standard Model (SM) particles, which are given by typically $\mathcal{O}(0.1-1)$ in terms of dimensionless couplings.
The reaction rate becomes smaller as the temperature decreases. The dark matter is decoupled from the thermal bath at around $m_\chi/T\sim25$, and thus the relic density is determined, where $m_\chi$ is the dark matter mass and $T$ is the temperature of the universe. 
In the latter case, the dark matter is never thermalized because interactions with SM particles are extremely small as $\mathcal{O}(10^{-11})$ 
for $m_\chi=\mathcal{O}(1)~\mathrm{GeV}$~\cite{Hall:2009bx}. 
The dark matter relic density is slowly generated via decays or scatterings of the thermal bath particles and is almost fixed at $m_\mathrm{th}/T\sim3$, 
where $m_\mathrm{th}$ is the mass of the particle in the thermal bath related to the dark matter production.\footnote{In fact, the first examples of the mechanism have been proposed in Refs.~\cite{Asaka:2005cn, Asaka:2006fs}.} 

These standard mechanisms for dark matter production may be altered by phase transitions (PTs) in the early universe. 
It has been claimed that the dark matter relic abundance is instantaneously determined by the first-order PTs if the masses of the particles relevant to the dark matter production 
largely change~\cite{Baker:2016xzo, Baker:2017zwx, Baker:2018vos}. 
As a result, the PTs may provide another solution for producing the observed dark matter relic abundance or may alleviate the tensions with the relevant constraints. 

If the PT is first-order, it produces stochastic gravitational waves (GW), whose spectra at the electroweak (EW) scale could be observed by various future space-based interferometers, such as the approved Laser Interferometer Space Antenna (LISA)~\cite{Caprini:2015zlo, LISA:2017pwj, Caprini:2019egz} and Deci-Hertz Interferometer Gravitational Wave Observatory (DECIGO)~\cite{Seto:2001qf, Kawamura:2011zz, Kawamura:2020pcg}, if the produced GW energy density is sufficiently large.
Hence, we could see it as a PT remnant affecting the dark matter production.
Besides, recently possibilities of multi-step PTs have been studied in the two Higgs doublet models~\cite{Fabian:2020hny, Aoki:2021oez, Benincasa:2022elt}. 
These articles have found that the multi-step PTs can be induced in certain parameter regions. 
This implies that the impact on the dark matter production may change depending on whether the PTs are one-step or multi-step. 

On the other hand, generating the small neutrino masses is an important subject. 
Although the neutrino oscillation experiments suggest that the neutrinos have small masses of $\mathcal{O}(0.1)~\mathrm{eV}$~\cite{deSalas:2020pgw}, they are exactly massless in the SM. 
The scotogenic model is one of the economical models accommodating the small neutrino masses and dark matter in the universe simultaneously~\cite{Ma:2006km}. 
In this model, the small neutrino masses are generated at the one-loop level, and the lightest $\mathbb{Z}_2$ odd particle (either a singlet fermion or a neutral inert scalar) can be a dark matter candidate, which is thermally produced by the freeze-out mechanism. 
However it is known that the fermionic dark matter candidate, the lightest right-handed neutrino, being similar to the bino-like dark matter 
in supersymmetric models~\cite{Ellis:1999mm, Arkani-Hamed:2006wnf}, has a tension with the theoretical and experimental constraints~\cite{Kubo:2006yx, Suematsu:2009ww, Schmidt:2012yg}. 
In particular, the constraint from the charged lepton flavor violation is severe, and we need some tuning among the parameters to avoid it~\cite{Toma:2013zsa}. 
The singlet fermionic dark matter can also be produced by the freeze-in mechanism~\cite{Molinaro:2014lfa, Hessler:2016kwm}. 
In this case, the second lightest $\mathbb{Z}_2$ odd particle becomes long-lived, which is expected to be probed at collider experiments~\cite{Hessler:2016kwm}.

In this work, we consider the effects of first-order PTs on the dark matter relic abundance in the scotogenic model.\footnote{See e.g.~Refs.~\cite{Borah:2020wut, Borah:2022cdx} for the previous studies of the PT effects in the (extended) scotogenic model.} 
Although some similar works have been done in Refs.~\cite{Bian:2018mkl, Bian:2018bxr}, our model setup is much simpler and can be regarded as a benchmark model for 
investigating effects of first-order PTs on dark matter relic abundance. 
In Section~\ref{sec:2}, we briefly review the model, theoretical and experimental constraints. 
In Section~\ref{sec:3}, we describe the effective potential at finite temperature for analyzing the PTs. 
We also perform a parameter search and give some benchmark parameter sets for the following calculation. 
Besides, gravitational waves generated by the first-order PTs are also examined.
Section~\ref{sec:4} is devoted to finding the effects of the first-order PTs on the dark matter relic abundance for the benchmark parameter sets.
Finally, we summarize our work in Section~\ref{sec:5}.

\section{The model}
\label{sec:2}
\subsection{The interactions and masses}
We consider the scotogenic model introducing three right-handed neutrinos $N_i~(i=1,2,3)$ with the mass $m_i$ and an inert doublet scalar $\eta$ with hypercharge $+1/2$ to the SM~\cite{Ma:2006km}. 
The $\mathbb{Z}_2$ parity is also imposed on the model such that the new particles are odd, whereas all the SM particles are even. 
The Lagrangian for the neutrino Yukawa coupling is given by
\begin{align}
\mathcal{L}_Y=-y_{i\alpha}\eta\overline{N_i}P_LL_\alpha+\mathrm{h.c.},
\label{eq:y}
\end{align}
where $L_\alpha=(\nu_\alpha,\ell_\alpha^-)^T$ is the SM lepton doublets with the flavor index $\alpha=e,\mu,\tau$. 
The scalar potential invariant under the symmetry is given by
\begin{align}
V_0=&~\mu_\Phi^2|\Phi|^2
+\mu_\eta^2|\eta|^2
+\lambda_1|\Phi|^4
+\lambda_2|\eta|^4 \nonumber\\
&+\lambda_3|\Phi|^2|\eta|^2
+\lambda_4|\eta^{\dag}\Phi|^2
+\frac{\lambda_5}{2}\left[\left(\eta^\dag\Phi\right)^2+\mathrm{h.c.}\right].
\label{eq:treepotential}
\end{align}
The scalar fields $\Phi$ and $\eta$ are parametrized as\footnote{It is known that the existence of the neutral CP-conserving vacuum indicates there are no deeper charge-breaking or CP-violating vacua at tree-level~\cite{Barroso:2013awa}. Hence, we do not consider such CP-violating vacua.}
\begin{align}
\Phi=\frac{1}{\sqrt{2}}\left(
\begin{array}{c}
 w^+\\
\phi_1+h+iz
\end{array}
\right),\qquad
\eta=\left(
\begin{array}{c}
 \eta^+\\
(\phi_2+H+iA)/\sqrt{2}
\end{array}
\right),
\end{align}
where $w^+$ and $z$ are the Nambu-Goldstone bosons absorbed by the $W$ boson and the $Z$ boson after the EW symmetry breaking. 
Note that only the neutral CP-even scalar field in $\Phi$ obtains the vacuum expectation value (VEV) at zero temperature, namely $\phi_1(T=0)\equiv v=246$ GeV.

The squared masses of the physical scalar particles at zero temperature are given by
\begin{align}
m_h^2&=2\lambda_1 v^2,\label{eq:h}\\
m_{\eta^\pm}^2&=\mu_\eta^2+\frac{\lambda_3}{2}v^2,\label{eq:etapm}\\
m_{H}^2&=\mu_\eta^2+\frac{\lambda_3+\lambda_4+\lambda_5}{2}v^2,\label{eq:H}\\
m_{A}^2&=\mu_\eta^2+\frac{\lambda_3+\lambda_4-\lambda_5}{2}v^2,\label{eq:A}
\end{align}
where the stationary condition $\mu_\Phi^2=-\lambda_1v^2$ is used, and $h$ corresponds to the SM Higgs boson, with the mass $m_h=125~\mathrm{GeV}$. 
From above equations, we can get
\begin{align}
\lambda_3 &= \frac{2}{v^2}(m_{\eta^\pm}^2 - \mu_\eta^2),\\
\lambda_4 &= \frac{m_H^2 + m_A^2 - 2m_{\eta^\pm}^2}{v^2},\\
\lambda_5 &= \frac{m_H^2 - m_A^2}{v^2}.
\end{align}
We take the masses of the inert scalar particles, $m_{\eta^\pm}, m_H$, and $m_{A}$, as input parameters instead of $\lambda_3, \lambda_4,$ and $\lambda_5$ in the scalar potential of Eq.~(\ref{eq:treepotential}). 
The other parameters $\mu_\eta^2$ and $\lambda_2$ remain as free parameters.

In this model, the neutrino masses are generated at the one-loop level, and the mass formula is given by
\begin{align}
(m_{\nu})_{\alpha\beta}=\sum_{i=1}^{3}\frac{y_{i\alpha}y_{i\beta}m_i}{2(4\pi)^2}
\left[
\frac{m_H^2}{m_H^2-m_i^2}\log\left(\frac{m_H^2}{m_i^2}\right)
-\frac{m_A^2}{m_A^2-m_i^2}\log\left(\frac{m_A^2}{m_i^2}\right)
\right]\equiv \left(y^{T}\Lambda y\right)_{\alpha\beta}.
\label{eq:nu-mass}
\end{align}
This mass matrix can be diagonalized by the Pontecorvo-Maki-Nakagawa-Sakata (PMNS) matrix $U_\mathrm{PMNS}$ as $U_\mathrm{PMNS}^Tm_{\nu}U_\mathrm{PMNS}=\hat{m}_{\nu}$ where $\hat{m}_{\nu}$ is the diagonalized neutrino mass matrix. 
The Yukawa matrix is given by the Casas-Ibarra parametrization~\cite{Casas:2001sr}
\begin{align}
y=\sqrt{\Lambda}^{-1}C\sqrt{\hat{m}_{\nu}}U_\mathrm{PMNS}^{\dag},
\label{eq:casas-ibarra}
\end{align}
where $C$ is a complex orthogonal matrix satisfying $C^TC=CC^T=\1_{3\times3}$. 
The neutrino mass eigenvalues and mixing angles consistent with the neutrino oscillation data can be reproduced as long as we choose this parametrization. 
In the following, we assume that the second and third right-handed neutrinos ($N_2$ and $N_3$) are much heavier than 
the first right-handed neutrino $N_1$ and the inert scalar particles. 
This assumption indicates that the neutrino Yukawa couplings naturally have the hierarchy $|y_{1\alpha}|\ll |y_{2\alpha}|, |y_{3\alpha}|$ via Eq.~(\ref{eq:casas-ibarra}).
Thus such hierarchical Yukawa couplings lead to fast decays of the heavier right-handed neutrinos into the lightest one, and we do not need to consider 
the effect of these particles on the dark matter relic abundance as will be seen later.

\subsection{The constraints}\label{sec:2.2}
We consider the theoretical constraints as follows.
First, the conditions for the tree level potential bounded from below are given by
\begin{align}
\lambda_1>0,~~ \lambda_2>0,~~ -2\sqrt{\lambda_1\lambda_2}<\lambda_3,~~ -2\sqrt{\lambda_1\lambda_2}<\lambda_3+\lambda_4-\lambda_5,
\end{align}
and the perturbativity conditions are given by 
\begin{align}
|\lambda_n|<4\pi\ (n=1,2,\cdots 5).
\end{align}
In addition to the above perturbative conditions, we consider the tree-level unitarity bound~\cite{Kanemura:1993hm, Akeroyd:2000wc, Arhrib:2012ia}.
Furthermore, we impose the condition that the EW vacuum is absolutely stable, namely it should not be metastable with a lifetime long enough compared to the age of the universe~\cite{Barroso:2013awa, Ivanov:2015nea}.
We numerically check that the EW vacuum is the global minimum in the region for $|\phi_i|\leq10$ TeV in the later part of the paper.

For the experimental constraints, the LEP precisely measured the decay widths of the $W$ and $Z$ bosons~\cite{Cao:2007rm} and does not allow a window for new decay channels into the inert scalar particles. 
Namely, the masses should satisfy 
\begin{equation}
m_H + m_{\eta^\pm} > m_W,~~ m_A + m_{\eta^\pm} > m_W,~~ m_H + m_A > m_Z,~~ 2 m_{\eta^\pm} > m_Z.
\end{equation}
Additionally neutral final states searches at the LEP~\cite{Lundstrom:2008ai} exclude the mass region
\begin{equation}
m_H < 80~\mathrm{GeV}~\cap~ m_A < 100~\mathrm{GeV}~\cap~ m_A - m_H > 8~\mathrm{GeV}.
\end{equation}
There is the lower bound for the charged inert scalar 
\begin{equation}
m_{\eta^\pm} > 70~\mathrm{GeV}.
\label{eq:metapm_const}
\end{equation}
This bound comes from the charged scalar pair production searches at LEP~\cite{Pierce:2007ut}. 
We also consider the constraints from the EW precision data, which can be avoided by assuming the mass degeneracy between the charged and neutral inert scalars ($m_{\eta^\pm}\approx m_{H}$ or $m_A$), 
because the custodial symmetry recovers in this limit~\cite{Haber:2010bw}.
Note that the neutral inert scalars do not contribute to the Higgs boson invisible decay because we set the neutral inert scalar masses such that any combination of those is smaller than the SM Higgs boson mass.

\section{The phase transitions}
\label{sec:3}
\subsection{Thermal effective potential}
The finite temperature effective potential at the one-loop $V^T$ is given by
\begin{align}
V^T = V_0 + V_{\rm CW} + V_{\rm CT} + V_1^T,
\label{eq:totalpotential}
\end{align}
where $V_0$ is the potential at the tree-level given by Eq.~(\ref{eq:treepotential}), $V_{\rm CW}$ is the one-loop Coleman-Weinberg potential~\cite{Quiros:1999jp}, 
$V_{\rm CT}$ is the counterterm potential, and $V_1^T$ is the one-loop effective potential at finite temperature~\cite{Dolan:1973qd}, respectively. 

The Coleman Weinberg potential can be written as~\cite{Quiros:1999jp}
\begin{align}
V_{\rm CW}(\phi_1, \phi_2)=\pm \frac{1}{64\pi^2}\sum_k
n_km_k^4(\phi_1, \phi_2)\left[\log\frac{m_k^2(\phi_1, \phi_2)}{\mu^2}-C_k
\right],
\label{eq:cwpotential}
\end{align}
where the index $k$ denotes particle species, $m_k$ and $n_k$ are the mass and degrees of freedom of a particle $k$. 
Specifically, the degrees of freedom are given by $n_k=2, 1, 1, 1, 6, 3, 2, 12, 12$, and 4 for $k=\eta^\pm, h, H, A, W, Z, \gamma, t, b$, and $\tau$, respectively.
Note that we include only the contributions from top and bottom quarks, and tau as fermions because the other
contributions are negligible due to its small Yukawa couplings.
The overall plus sign in Eq.~(\ref{eq:cwpotential}) is for the boson contributions, while the minus sign is for the fermion contributions. 
The renormalization scale denoted by $\mu$ is fixed to be $\mu=v$.
Because we choose the $\overline{{\rm MS}}$ scheme for the renormalization, the coefficients $C_k$ are $1/2$ for the transverse gauge bosons but $3/2$ for the other particle species.
In general, the quantum corrections shift the position of the EW vacuum and the masses of the scalar fields from the original ones. 
The counterterm $V_{\rm CT}$ is determined, such that these quantities are fixed at the tree-level,
\begin{align}
\left.\frac{\partial V_{\rm CT}(\phi_1, \phi_2)}{\partial\phi_i}\right|_{(\phi_1,\phi_2)=(v, 0)}
&=-\left.\frac{\partial V_{\rm CW}(\phi_1, \phi_2)}{\partial\phi_i}\right|_{(\phi_1,\phi_2)=(v, 0)}\equiv -V_{i},\\
\left.\frac{\partial^2 V_{\rm CT}(\phi_1, \phi_2)}{\partial\phi_i\partial\phi_j}\right|_{(\phi_1,\phi_2)=(v, 0)}
&=-\left.\frac{\partial^2 V_{\rm CW}(\phi_1, \phi_2)}{\partial\phi_i\partial\phi_j}\right|_{(\phi_1,\phi_2)=(v, 0)} \equiv -V_{ij}\ \ (i, j=1, 2).
\label{eq:CTconditions}
\end{align}
We set $V_{\rm CT}$ following Ref.~\cite{Fabian:2020hny}, as
\begin{align}
V_{\rm CT}=\delta \mu_\Phi^2\phi_1^2+\delta \mu_\eta^2\phi_2^2+\delta\lambda_1\phi_1^4,
\label{eq:VCT}
\end{align}
with 
\begin{equation}
\begin{aligned}
&\delta \mu_\Phi^2=-\frac{3}{4v}V_1+\frac{1}{4}V_{11},\\
&\delta \mu_\eta^2=-\frac{1}{2}V_{22},\\
&\delta \lambda_1=\frac{1}{8v^3}(V_1-v V_{11}).
\end{aligned}
\end{equation}
When calculating the coefficients $\delta \mu_\Phi^2, \delta \mu_\eta^2$, and $\delta\lambda_1$ numerically, infrared log divergences arise in the second derivatives of $V_{\rm CW}$ for the Nambu-Goldstone boson contributions. 
These divergences are removed by introducing a cut-off for the Nambu-Goldstone boson masses. Here we take the cut-off at $m_h$~\cite{Cline:2011mm}.

The one-loop effective potential at finite temperature is given by~\cite{Dolan:1973qd}
\begin{align}
V_1^T(\phi_1, \phi_2)=
\pm\frac{T^4}{2\pi^2}\sum_k
\int_0^\infty dx\ x^2\ln\left[1\mp \exp\left(-\sqrt{x^2+\frac{m_k^2(\phi_1, \phi_2)}{T^2}}\right)\right],
\label{eq:thermalpotential}
\end{align}
with the temperature $T$. The upper (lower) sign is for bosonic (fermionic) contributions.

We take into account the corrections from the ring (Sazae-san) diagrams to avoid infrared divergences arising in loop expansions~\cite{Parwani:1991gq, Arnold:1992rz}.
For this purpose, we apply the Parwani resummation method~\cite{Parwani:1991gq}, which is suitable for evaluating multi-step PTs because this method does not contain the high-temperature expansion~\cite{Aoki:2021oez}.\footnote{The state-of-the-art computations of the effective potential including higher order terms of perturbation have been studied in Refs.~\cite{Croon:2020cgk, Schicho:2022wty, Ekstedt:2022bff, Biondini:2022ggt}.} 
We can perform this method by inserting masses with thermal corrections to $m_k^2$ in $V_1^T$.
The thermal masses used for this method are described in the following subsection.

\subsection{Thermal masses}
\label{sec:thermalmass}
The inert scalar field $\eta$ may temporarily have a VEV at finite temperature. 
Then, it should vanish at zero temperature not to spontaneously break the $\mathbb{Z}_2$ parity. 
Thus, the mass matrices for the inert scalar particles at finite temperature can have a different shape from those at zero temperature. 
The mass matrices for the charged scalars $(w^+,\eta^+)$, the CP-even scalars $(h, H)$ and the CP-odd scalars $(z, A)$ are field-dependent and given by
\begin{align}
\mathcal{M}_{\rm charge}^2=&~\frac{1}{2}
\left(
\begin{array}{cc}
2\mu_\Phi^2+2\lambda_1\phi_1^2+\lambda_3\phi_2^2+c_1T^2 & (\lambda_4+\lambda_5)\phi_1 \phi_2 \\
(\lambda_4+\lambda_5)\phi_1 \phi_2 & 2\mu_\eta^2+2\lambda_2\phi_2^2+\lambda_3\phi_1^2+c_2T^2
\end{array}
\right),\nonumber\\
\mathcal{M}_{\rm odd}^2=&~\frac{1}{2}
\left(
\begin{array}{cc}
2\mu_\Phi^2+2\lambda_1\phi_1^2+\bar{\lambda}_{345}\phi_2^2+c_1T^2
& 2\lambda_5\phi_1 \phi_2 \\
2\lambda_5\phi_1 \phi_2 &
2\mu_\eta^2+2\lambda_2\phi_2^2+\bar{\lambda}_{345}\phi_1^2+c_2T^2
\end{array}
\right),\\
\mathcal{M}_{\rm even}^2=&~\frac{1}{2}
\left(
\begin{array}{cc}
2\mu_\Phi^2+6\lambda_1\phi_1^2+\lambda_{345}\phi_2^2+c_1T^2 &
2\lambda_{345}\phi_1 \phi_2 \\
2\lambda_{345}\phi_1 \phi_2 &
2\mu_\eta^2+6\lambda_2\phi_2^2+\lambda_{345}\phi_1^2+c_2T^2 
\end{array}
\right),\nonumber
\label{eq:bosonmassmatrices1}
\end{align}
where the coupling $\bar{\lambda}_{345}$ is defined as $\bar{\lambda}_{345}\equiv \lambda_3 + \lambda_4 - \lambda_5$, and the coefficients $c_1$ and $c_2$ are given by
\begin{align}
&c_1=\frac{1}{8}g^2+\frac{1}{16}(g^2+g'^2)+\frac{1}{2}\lambda_1+\frac{1}{6}\lambda_3+\frac{1}{12}\lambda_4,\\
&c_2=\frac{1}{8}g^2+\frac{1}{16}(g^2+g'^2)+\frac{1}{2}\lambda_2+\frac{1}{6}\lambda_3+\frac{1}{12}\lambda_4 +\frac{1}{4}y_t^2+\frac{1}{4}y_b^2+\frac{1}{12}y_\tau^2,
\label{eq:thermalcorrection}
\end{align}
with the $SU(2)_L$ and $U(1)_Y$ gauge couplings $g$ and $g'$, and the Yukawa couplings $y_t,y_b$ and $y_\tau$ for top, bottom, and tau, respectively. 
We regard the mass eigenvalues obtained by diagonalizing the above mass matrices as the physical masses with thermal effects.

The gauge bosons also get thermal corrections. 
In fact, only the longitudinal modes get thermal corrections, whereas the transverse modes do not. 
The masses for the transverse modes of the $W$ boson, $Z$ boson, and photon are written as
\begin{align}
m_{W_T}^2 =&~ \frac{g^2}{4}(\phi_1^2+\phi_2^2),\\
m_{Z_T}^2=&~\frac{1}{4}(g^2+g'^2)(\phi_1^2+\phi_2^2),\\
m_{\gamma_T}^2=&~0.
\label{eq:gaugethermaleffect}
\end{align}
On the other hand, the $W$ boson mass for the longitudinal mode with thermal corrections is given by~\cite{Blinov:2015vma}
\begin{equation}
m_{W_L}^2=\frac{g^2}{4}(\phi_1^2+\phi_2^2) +2g^2T^2.
\end{equation}
For the longitudinal $Z$ boson and photon, the masses are obtained by diagonalizing the following mass matrix
\begin{align}
\frac{1}{4}(\phi_1^2+\phi_2^2)
\left(
\begin{array}{cc}
g^2&-gg'\\
-gg'&g'^2\\
\end{array}
\right)
+
\left(
\begin{array}{cc}
2g^2T^2&0\\
0&2g'^2T^2\\
\end{array}
\right)\,.
\end{align}
Then, we obtain the mass eigenvalues as 
\begin{align}
m_{Z_L}^2=&~\frac{1}{8}(g^2+g'^2)(\phi_1^2+\phi_2^2+8T^2)
+\Delta,\\
m_{\gamma_L}^2=&~\frac{1}{8}(g^2+g'^2)(\phi_1^2+\phi_2^2+8T^2)
-\Delta,
\label{eq:gaugethermaleffect}
\end{align}
where $\Delta$ is given by
\begin{align}
\Delta=\sqrt{\frac{1}{64}(g^2+g'^2)^2(\phi_1^2+\phi_2^2+8T^2)^2-g^2g'^2T^2(\phi_1^2+\phi_2^2+4T^2)}.
\end{align}

\subsection{Parameter search and benchmark parameter sets}

\begin{figure}[t]
\begin{center}
 \includegraphics[width=5cm]{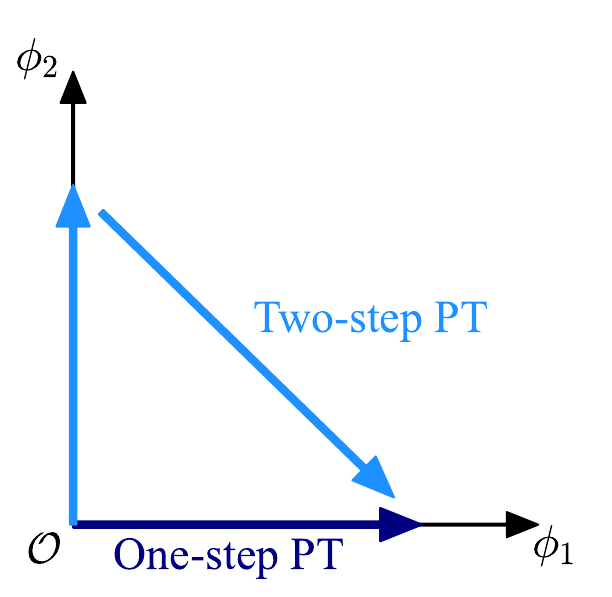}
\caption{Schematic picture of one-step and two-step PTs.}
\label{fig:schematic}
\end{center} 
\end{figure}

In general, various PT patterns are possible. 
In the following analyses, we consider only the patterns shown in Fig.~\ref{fig:schematic}.
The one-step PT denotes the transition occurring only once from the origin $(\phi_1, \phi_2)=(0,0)$ to the EW vacuum $(v,0)$. 
Besides, the two-step PT is referred to the transition going to the EW vacuum via two steps: $(\phi_1,\phi_2)=(0,0)\to(0,v^\prime)\to(v,0)$.
The 2nd PT of the two-step must be first-order because the potential barrier (saddle point of the potential) exists between the two minima.
There is another possibility of two-step PTs: $(\phi_1,\phi_2)=(0,0)\to(v^{\prime\prime},v^\prime)\to(v,0)$. 
However this transition hardly occurs in the scotogenic model because the quadratic cross term $\Phi^\dag\eta$ in the scalar potential is forbidden by the $\mathbb{Z}_2$ parity. 

To reveal the impact of PTs on the dark matter relic density, we perform a parameter search using CosmoTransitions for evaluating PTs~\cite{Wainwright:2011kj}. 
The independent parameters are randomly selected $5\times10^6$ times in total from the intervals:
\begin{align*}
&
70~\mathrm{GeV}< m_{\eta^\pm}=m_A\leq1000~\mathrm{GeV},
\quad
55~\mathrm{GeV}\leq m_H\leq 1000~\mathrm{GeV},\\
&
0\leq \lambda_2 \leq 4\pi,
\quad
-(1000~\mathrm{GeV})^2\leq \mu_{\eta}^2\leq (1000~\mathrm{GeV})^2,
\end{align*}
where the lower bound on $m_{\eta^\pm}$ follows Eq.~(\ref{eq:metapm_const}).
We assume that the inert scalar masses for $\eta^\pm$ and $A$ are degenerate $(m_{\eta^\pm}=m_A)$ so that the constraint of the EW precision measurements is easily satisfied as referred to in Sec.~\ref{sec:2.2}.
Tab.~\ref{tb:eachparameterregion} shows the minimum and maximum values of each parameter obtained by our parameter search and the number of parameter sets (last column) calculated, considering the relevant constraints. 
The fourth and fifth rows with the symbol $^*$ denote the parameter sets satisfying the conditions $m_\eta/T_n>25$ and $\phi_2(T_{n2})/T_{n2}>1$, respectively. 
Here $m_\eta$ means the smallest mass of the inert scalar boson and
the nucleation temperature $T_n$ is determined by the condition of one bubble nucleation per Hubble radius: $S_3(T_n)/T_n\simeq140$ with the $O(3)$ symmetric action $S_3$. 
For the two-step PT, we define the nucleation temperature of the first step as $T_{n1}$ and that of the second step $T_{n2}$.
Note that the one-step$^*$ PT is necessarily first-order because the conditions $m_\eta/T_n>25$ demands $T_n$ to be very low temperature.

\begin{figure}[t]
\begin{center}
\includegraphics[width=7.5cm]{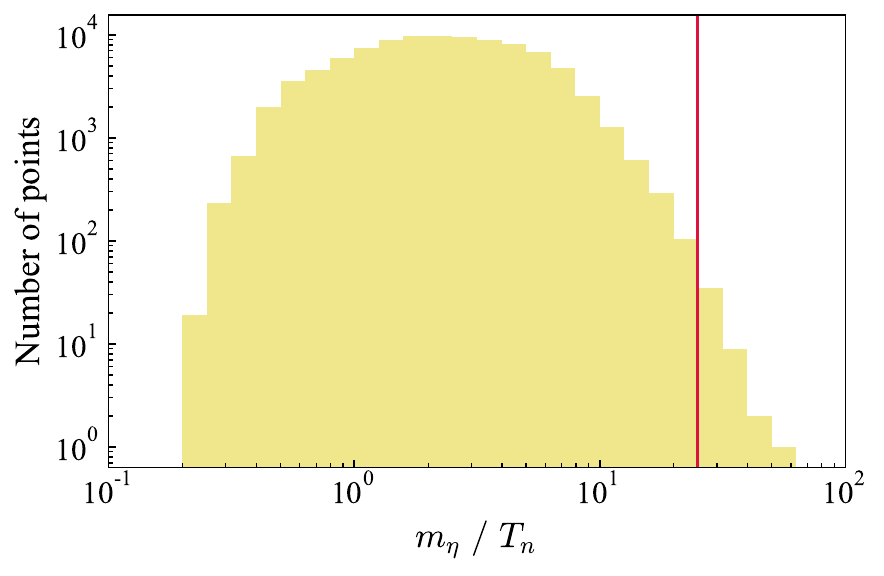}
\includegraphics[width=7.5cm]{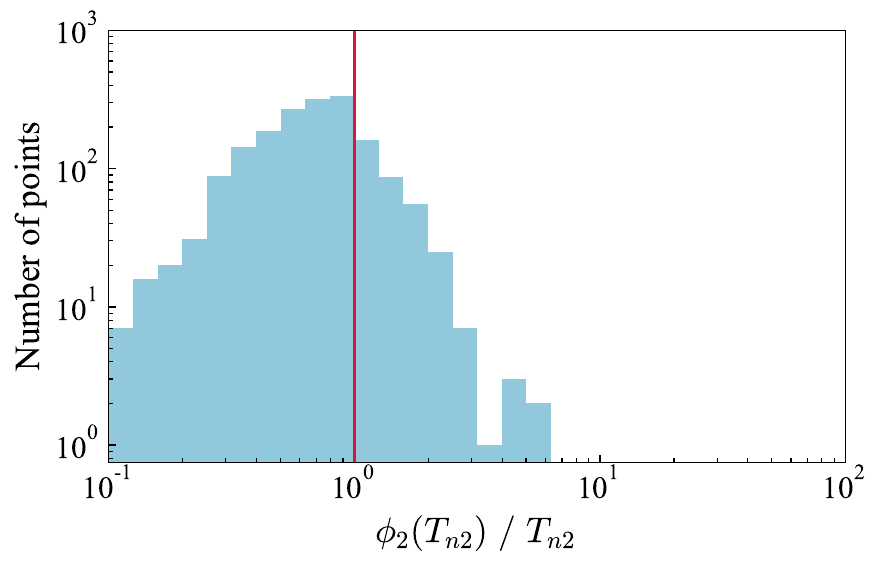}
\caption{Distribution of the number of parameter sets as a function of $m_\eta/T_n$ for the one-step PTs (left) and $\phi_2(T_{n2})/T_{n2}$ for the two-step (right) PTs, respectively.
The red vertical lines show the thresholds $m_\eta/T_n=25$ and $\phi_2(T_{n2})/T_{n2}=1$. 
}
\label{fig:numberanalysis}
\end{center}
\end{figure}

Only the parameter sets with the symbol $^*$ can potentially influence the calculation of dark matter relic abundance. One can find that the parameter region affected by the PTs is limited. 
Therefore the previous studies on dark matter in the scotogenic model remain correct except for some specific parameter regions.
However we emphasize that $19\%~(=333/1758)$ of two-step PTs can have an impact on the dark matter relic density, which is not small. 
Fig.~\ref{fig:numberanalysis} shows the distribution of the number of parameter sets as a function of $m_{\eta}/T_n$ for the one-step PTs and $\phi_2(T_{n2})/T_{n2}$ for the two-step PTs.
The red vertical lines show the thresholds $m_\eta/T_n=25$ and $\phi_2(T_{n2})/T_{n2}=1$. 
The parameter sets with the symbol $^*$ in Tab.~\ref{tb:eachparameterregion} correspond to results on the right side of the red line in Fig.~\ref{fig:numberanalysis}.

\begin{table}[t]
\caption{
Minimum and maximum values of the independent parameters for each PT obtained from the parameter search ($m_A=m_{\eta^\pm}$). 
The last column represents the number of parameter sets calculated considering the relevant constraints. 
The symbol $^*$ in the fourth and fifth rows denotes the PTs which can potentially influence the calculation of dark matter relic abundance.
}
\centering
\begin{tabular}{|c||c|c|c|c|c|}\hline
PT type & $m_{\eta^\pm}~[\mathrm{GeV}]$ & $m_H~[\mathrm{GeV}]$ & $\lambda_2$ & $\mu_{\eta}^2~[\mathrm{GeV}^2]$ & Results $\#$ \\
\hhline{|=#=|=|=|=|=|}
All & ($70$, $1000$] & [55, $1000$] & [$0$, $4.2$] & [$-(311)^2$, $(992)^2$] & 97684 (100$\%$)\\
One-step & ($70$, $1000$] & [55, $1000$] & [$0$, $4.2$] & [$-(311)^2$, $(992)^2$] & 95926 (98.2$\%$)\\
Two-step & (70, 440] & [$55$, $660$] & [$0.13$, $4.1$] & [$-(214)^2$, $(298)^2$] & 1758 (1.80$\%$)\\ \hline
One-step$^*$ & [698, 996] & [$575$, $987$] & [$0.18$, $2.0$] & [$(508)^2$, $(801)^2$] & 47 (0.0481$\%$)\\
Two-step$^*$ & [70, 434] & [$55$, $570$] & [$0.13$, $4.1$] & [$-(214)^2$, $(189)^2$] & 333 (0.341$\%$)\\\hline
\end{tabular}
\label{tb:eachparameterregion}
\end{table}

\begin{table}[t]
\begin{center}
\caption{Benchmark parameters leading first-order PTs where the inert scalar masses $m_{\eta^\pm},m_H$ and $m_A$ represent the values at zero temperature ($m_{\eta^\pm}=m_A$).}
\label{tab:1}
 \begin{tabular}{|c||c|c|c|c|c|}\hline
  & $m_{\eta^\pm}~[\mathrm{GeV}]$ & $m_H~[\mathrm{GeV}]$ & $\lambda_2$ & $\mu_\eta^2~[\mathrm{GeV}^2]$ & PT type\\
\hhline{|=#=|=|=|=|=|}
BM1 & $682$ & $729$ & $1.64$ & $(479)^2$ & one-step$^*$\\
BM2 & $958$ & $943$ & $0.672$ & $(769)^2$ & one-step$^*$\\\hline
BM3 & $136$ & $205$ & $1.14$ & $-(136)^2$ & two-step$^*$\\
BM4 & $151$ & $130$ & $1.31$ & $-(147)^2$ & two-step$^*$\\\hline
 \end{tabular}
\label{tab:BM}
\end{center}
\end{table}

\begin{figure}[t]
 \begin{center}
  \includegraphics[width=7.5cm]{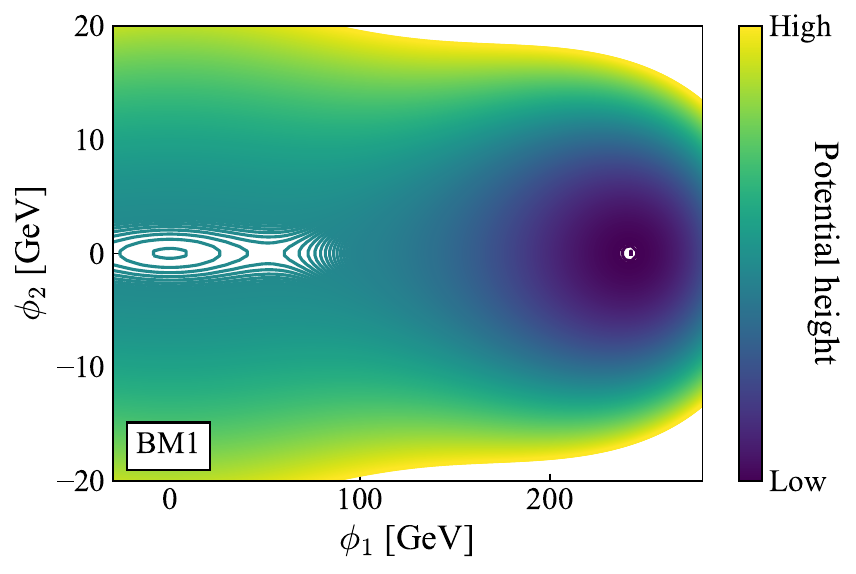}
  \includegraphics[width=7.5cm]{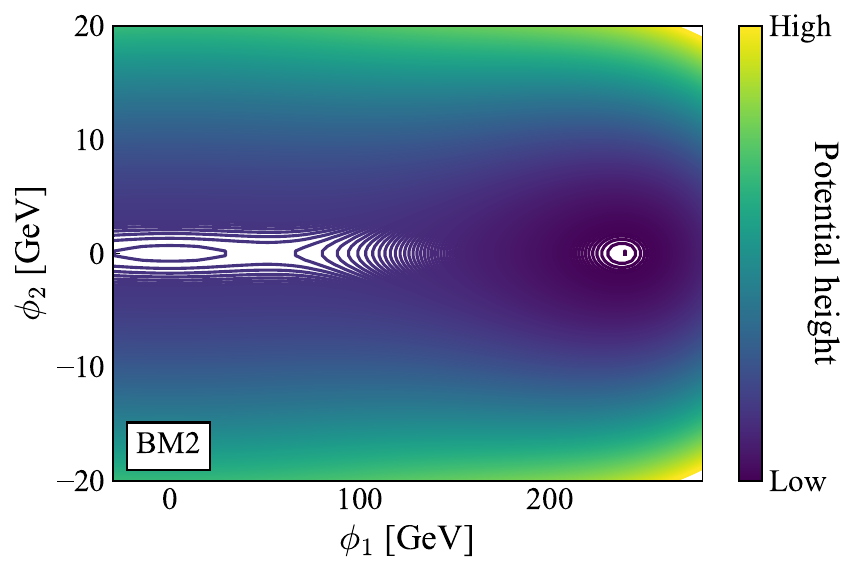}
\caption{Contours of the finite temperature effective potential $V^T$ at $T=T_n$ for the BM1 (left) and BM2 (right). 
The nucleation temperature is determined as $T_n=23~\mathrm{GeV}$ for BM1 and $25~\mathrm{GeV}$ for BM2.
The first-order PT occurs from the origin to the EW vacuum in both benchmark parameter sets.
}
\label{fig:contour1}
 \end{center}
\end{figure}

To see the impact of PTs on the dark matter relic density specifically, we choose four benchmark parameter sets in Tab.~\ref{tab:BM}.
All constraints in Sec.~\ref{sec:2.2} are satisfied at the benchmark parameter sets.
Fig.~\ref{fig:contour1} shows the contours of the thermal effective potential at $T=T_n$ for BM1 and BM2 
where the nucleation temperature is determined as $T_n=23~\mathrm{GeV}$ for BM1 and $T_n=25~\mathrm{GeV}$ for BM2.\footnote{
One may consider $\sqrt{\phi_1^2+\phi_2^2}/T$ at critical temperatures are too large as $10$~\cite{Basler:2016obg} because of such a low $T_n$.
However, those values are not so sizable ($4$--$4.8$) for BM1 and BM2.
}
For both benchmark parameter sets, the first-order PTs occur from the origin to the EW vacuum. 

\begin{figure}[t]
 \begin{center}
  \includegraphics[width=7.5cm]{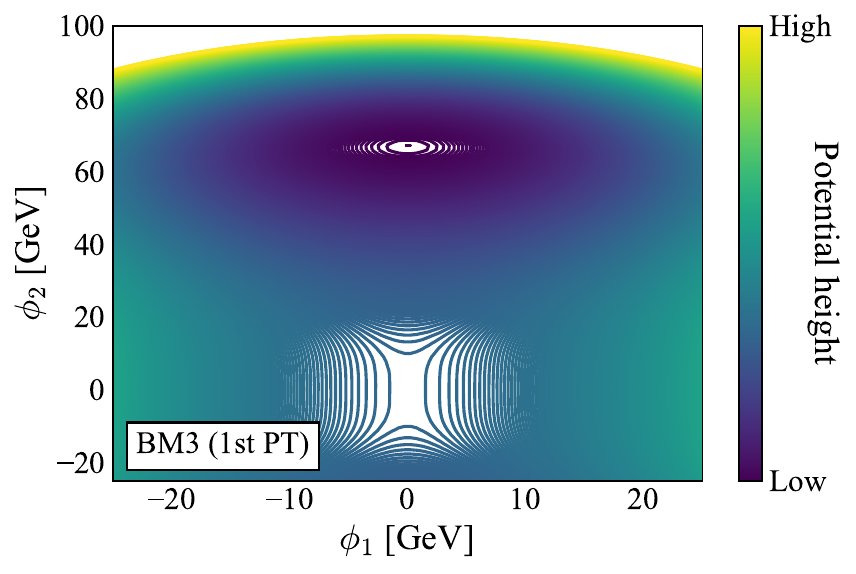}
  \includegraphics[width=7.5cm]{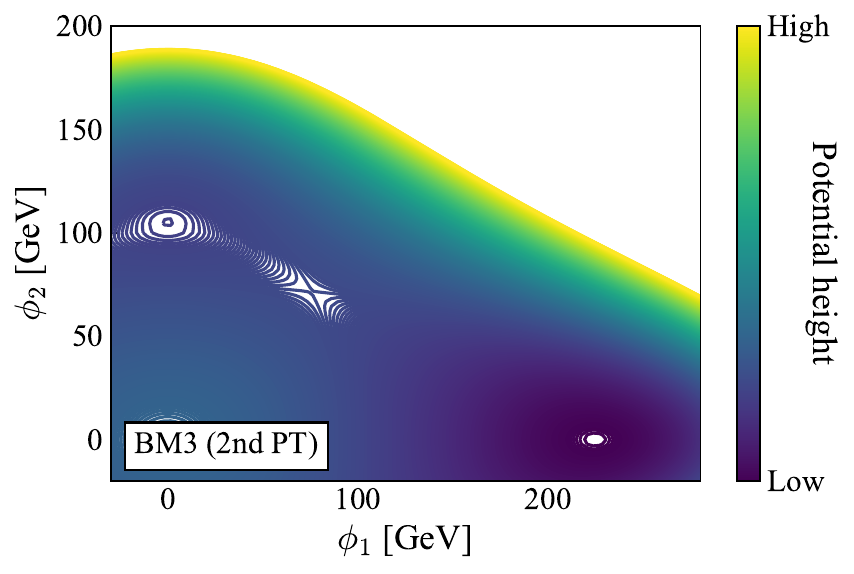}\\
  \includegraphics[width=7.5cm]{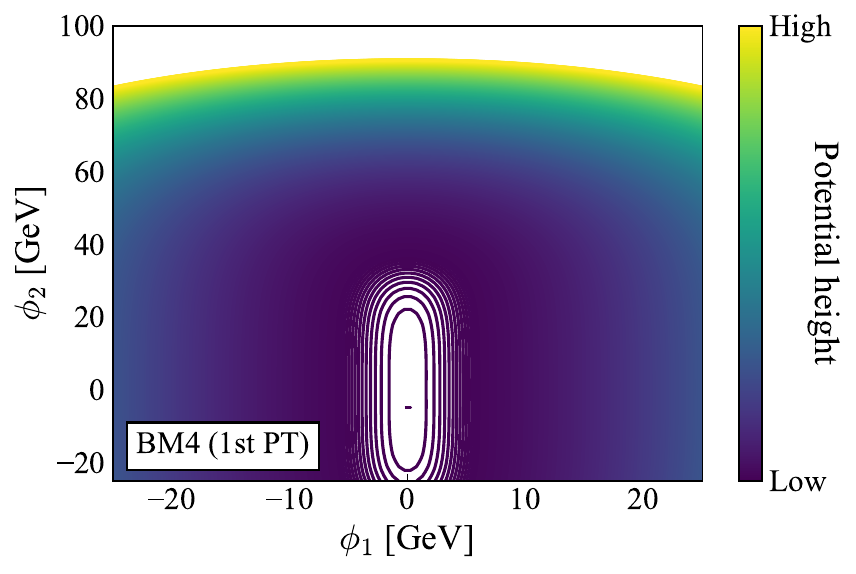}
  \includegraphics[width=7.5cm]{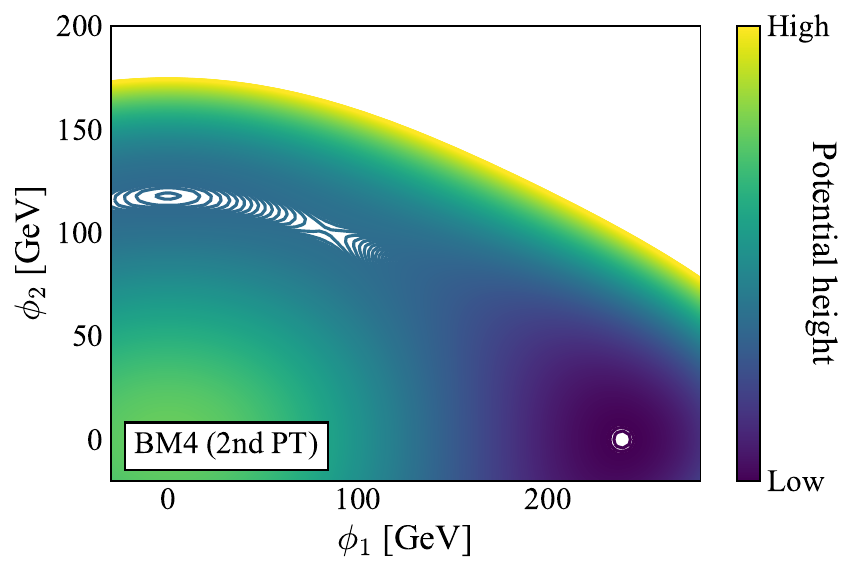}
\caption{Contours of the finite temperature effective potential for BM3 (top) and BM4 (bottom) at the nucleation temperatures, 
The nucleation temperatures are determined as $T_{n1}=148~\mathrm{GeV}$ (top left) and $T_{n2}=86~\mathrm{GeV}$ (top right) for BM3 
while $T_{n1}=184~\mathrm{GeV}$ (bottom left) and $T_{n2}=66~\mathrm{GeV}$ bottom (right) for BM4. 
The first PTs in the left panels occur from the origin to $(0,\phi_2(T_{n1}))$ along the $\phi_2$ axis, whereas the second PTs in the right panels occur from $(0,\phi_2(T_{n2}))$ to the EW vacuum.
}
\label{fig:contour2}
 \end{center}
\end{figure}

The two-step PTs occur for BM3 and BM4, unlike BM1 and BM2. 
Fig.~\ref{fig:contour2} represents the contours of the finite temperature effective potential for BM3 (top) and BM4 (bottom) at $T=T_{n1}$ (left) and $T_{n2}$ (right) 
where $T_{n1}$ and $T_{n2}$ are the nucleation temperatures for the first and second step PTs, 
respectively.\footnote{The nucleation temperature cannot be defined for the first PT for BM4 because the bubbles are not produced for the second-order PT.
Nevertheless, we simply regard $T_{n1}$ for BM4 as the temperature where the second-order PT occurs in this paper.
} 
The nucleation temperatures for BM3 are numerically determined as $T_{n1}=148~\mathrm{GeV}$ and $T_{n2}=86~\mathrm{GeV}$ 
while these are $T_{n1}=184~\mathrm{GeV}$ and $T_{n2}=66~\mathrm{GeV}$ for BM4. 
We confirmed that the numerical determination of the nucleation temperatures with CosmoTransitions is stable enough near the BM parameter sets.

\subsection{Gravitational waves}
GWs are generated by the first-order PTs. 
The generated GW energy density is described by the two parameters $\alpha$ and $\tilde{\beta}$, which are defined by
\begin{align}
\alpha \equiv& \left.\frac{\epsilon(T)}{\rho_{\rm rad}(T)}\right|_{T=T_n}\,,\label{eq:alpha}\\
\tilde\beta \equiv& \left.\left[T\frac{d}{dT}\left(\frac{S_3(T)}{T}\right)\right]\right|_{T=T_n}\, ,\label{eq:beta}
\end{align}
at the nucleation temperature $T_n$. 
The parameter $\alpha$ represents the PT strength and $\tilde \beta$ is equal to $\beta/H(T=T_n)$ where $\beta$ describes the inverse time duration of the PT and $H(T=T_n)$ is the Hubble rate at $T=T_n$.
In the case of two-step PTs, the above two parameters are obtained at each nucleation temperature $T_{n1}$ and $T_{n2}$. 
In Eq.~(\ref{eq:alpha}), $\epsilon (T)$ is the latent heat, and $\rho_{\rm rad}(T)$ is the radiation density given by $\rho_{\rm rad}(T)=\pi^2g_*T^4/30$ with the effective relativistic degrees of freedom in the symmetric phase $g_*$. 

There are three contributions to the GW energy density, which are given by
\begin{align}
\Omega_\text{GW} h^2 =
\Omega_\varphi h^2 +
\Omega_\text{sw} h^2 + 
\Omega_\text{turb} h^2\,,
\label{GW-Omega}
\end{align}
where $h$ denotes the dimensionless Hubble rate,
$\Omega_\varphi$ is the scalar field contribution from bubble wall collisions~\cite{Kosowsky:1991ua, Kosowsky:1992rz, Kosowsky:1992vn, Kamionkowski:1993fg, Caprini:2007xq,Huber:2008hg},
$\Omega_{\rm sw}$ is the sound wave contribution surrounding the bubble walls~\cite{Hindmarsh:2013xza, Hindmarsh:2015qta},
and $\Omega_{\rm turb}$ is the contribution from magnetohydrodynamic turbulence 
in plasma~\cite{Caprini:2006jb, Kahniashvili:2008pf, Kahniashvili:2008pe, Kahniashvili:2009mf, Caprini:2009yp, Binetruy:2012ze}. 
Numerical simulations and analytic estimation formulate each contribution to the GW energy density as follows~\cite{Huber:2008hg, Hindmarsh:2015qta, Caprini:2009yp, Binetruy:2012ze}, 
\begin{align}
\Omega_{\varphi}h^2=&~
1.67\times 10^{-5}
\tilde\beta^{-2}
\left(\frac{\kappa_{\varphi} \alpha}{1+\alpha}\right)^2
\left(\frac{100}{g_{*}}\right)^{\frac{1}{3}}\left(\frac{0.11v^3_{w}}{0.42+v_w^2 }\right)\frac{3.8(f/f_{\varphi})^{2.8}}{1+2.8(f/f_{\varphi})^{3.8}},
\label{eq:gw1}\\
%%%
\Omega_{\mathrm{sw}}h^2=&~
2.65\times 10^{-6}
\tilde\beta^{-1}
\left(\frac{\kappa_\text{sw} \alpha}{1+\alpha}\right)^2\left(\frac{100}{g_{*}}\right)^{\frac{1}{3}} v_{w}(f/f_{\mathrm{sw}})^3\left(\frac{7}{4+3(f/f_{\mathrm{sw}})^2}\right)^{\frac{7}{2}},
\label{eq:gw2}\\
%%%
\Omega_{\mathrm{turb}}h^2=&~
3.35\times 10^{-4}
\tilde\beta^{-1}
\left(\frac{\kappa_\text{turb}\alpha}{1+\alpha}\right)^{\frac{3}{2}}\left(\frac{100}{g_{*}}\right)^{\frac{1}{3}}
v_{w}\frac{(f/f_{\mathrm{turb}})^3}{[
1+(f/f_{\mathrm{turb}})]^{\frac{11}{3}}(1+8\pi f/h_n)},
\label{eq:gw3}
\end{align}
where $v_w$ is the bubble wall velocity taken to be $v_w=1$ for simplicity. 
The additional parameters $\kappa_\varphi$, $\kappa_{\mathrm{sw}}$, and $\kappa_{\rm turb}$ are the fraction of vacuum energy converted into 
gradient energy of the scalar field, the bulk motion of the fluid, and magnetohydrodynamic turbulence, respectively. 
These are estimated as~\cite{Kamionkowski:1993fg, Espinosa:2010hh, Hindmarsh:2015qta}
\begin{align}
\kappa_\varphi \approx&~ \frac{1}{1+0.715\alpha}\left(0.715\alpha+\frac{4}{27}\sqrt{\frac{3\alpha}{2}}\right),\\
\kappa_{\mathrm{sw}} \approx&~ \frac{\alpha}{0.73+0.083\sqrt{\alpha}+\alpha},\\
\kappa_{\rm turb} \approx&~ 0.1 \kappa_{\mathrm{sw}}.
\end{align}
In Eqs.~(\ref{eq:gw1})--(\ref{eq:gw3}), $f$ is the GW frequency at present, and 
the peak frequencies $f_\varphi$, $f_\mathrm{sw}$ and $f_\mathrm{turb}$ are given by~\cite{Huber:2008hg, Hindmarsh:2015qta, Caprini:2009yp, Binetruy:2012ze}
\begin{align}
f_{\varphi} =&~ 1.65 \times 10^{-5} \mathrm{Hz}~
\tilde\beta
\left(\frac{0.62}{1.8-0.1 v_w+v^2_w }\right)\left( \frac{T_n}{100~\mathrm{GeV}}\right)\left( \frac{g_{*}}{100}\right)^{\frac{1}{6}},\\
%%%
 f_{\mathrm{sw}} =&~ 1.9 \times 10^{-5} \mathrm{Hz}~
v_w ^{-1}
\tilde\beta
 \left( \frac{T_n}{100~\mathrm{GeV}}\right)\left( \frac{g_{*}}{100}\right)^{\frac{1}{6}}\Upsilon,\\
%%%
f_{\mathrm{turb}} =&~ 2.7 \times 10^{-5} \mathrm{Hz}~
v_w ^{-1}
\tilde\beta
 \left( \frac{T_n}{100~\mathrm{GeV}}\right)\left( \frac{g_{*}}{100}\right)^{\frac{1}{6}}\,,
\end{align}
where $\Upsilon$ is the suppression factor due to the finite lifetime of sound waves, which is described by~\cite{Guo:2020grp}
\begin{equation}
\Upsilon=1-\frac{1}{\sqrt{1+2\tau_{sw}H_c}}\,,
\end{equation}
with 
\begin{align}
 \tau_{sw}H_c\sim (8\pi)^{1/3}v_w\tilde\beta\sqrt{\frac{4(1+\alpha)}{3\alpha \kappa_\mathrm{sw}}}.
\end{align}
Finally, $h_n$ in Eq.~(\ref{eq:gw3}) is the value of the inverse Hubble time at $T=T_n$ redshifted today, which is given by
\begin{align}
h_n = 1.65 \times 10^{-5} \mathrm{Hz}~\left( \frac{T_n}{100~\mathrm{GeV}}\right)\left( \frac{g_{*}}{100}\right)^{1/6}\,.
\end{align}

\begin{figure}[t]
 \begin{center}
  \includegraphics[width=8.5cm]{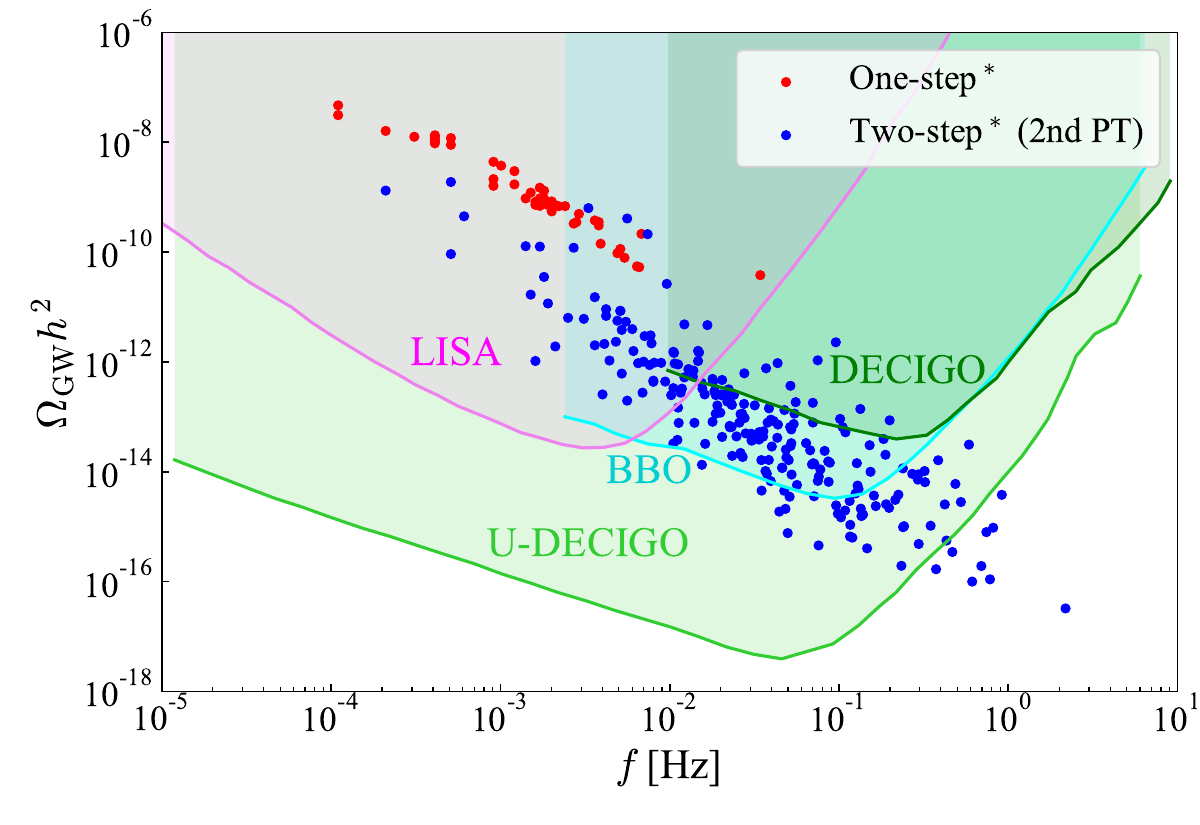}
\caption{Peaks of the GW energy density fraction as a function of the frequency for the one-step$^*$ PT (red) and the 2nd PT of the two-step$^*$ (blue). The colored regions are observable areas by future space-based interferometers.}
\label{fig:GWall}
 \end{center}
\end{figure}

We show the peaks of the GW density fraction as a function of the frequency $f$ for the one-step$^*$ PT (red) and the 2nd PT of the two-step$^*$ (blue) in Fig.~\ref{fig:GWall}.
The colored regions in Fig.~\ref{fig:GWall} represent the observable areas in the future interferometers, such as LISA~\cite{Caprini:2015zlo, LISA:2017pwj, Caprini:2019egz}, DECIGO~\cite{Seto:2001qf, Kawamura:2011zz, Kawamura:2020pcg}, Big Bang Observer (BBO)~\cite{Corbin:2005ny}, and Ultimate DECIGO (U-DECIGO)~\cite{Kudoh:2005as}.
We can see all peaks for the one-step$^*$ PT represent $\Omega_\text{GW} h^2\gtrsim10^{-10}$ with $10^{-4}<f<10^{-1}$ and can be detected by LISA.
This is because the condition $m_\eta/T_n>25$ forces the nucleate temperature very low and makes the PT first-order with the huge latent heat. 
On the other hand, the most of peaks for 2nd PT of the two-step$^*$ PT locate the upper side of the detectable area by U-DECIGO.
Although the condition $\phi_2(T_{n2})/T_{n2}>1$ is not so strict as $m_\eta/T_n>25$, it also forces $T_{n2}$ rather low.
Therefore, future observations may clarify the two-step$^*$ PT scenario.

\begin{figure}[t]
 \begin{center}
  \includegraphics[width=8.5cm]{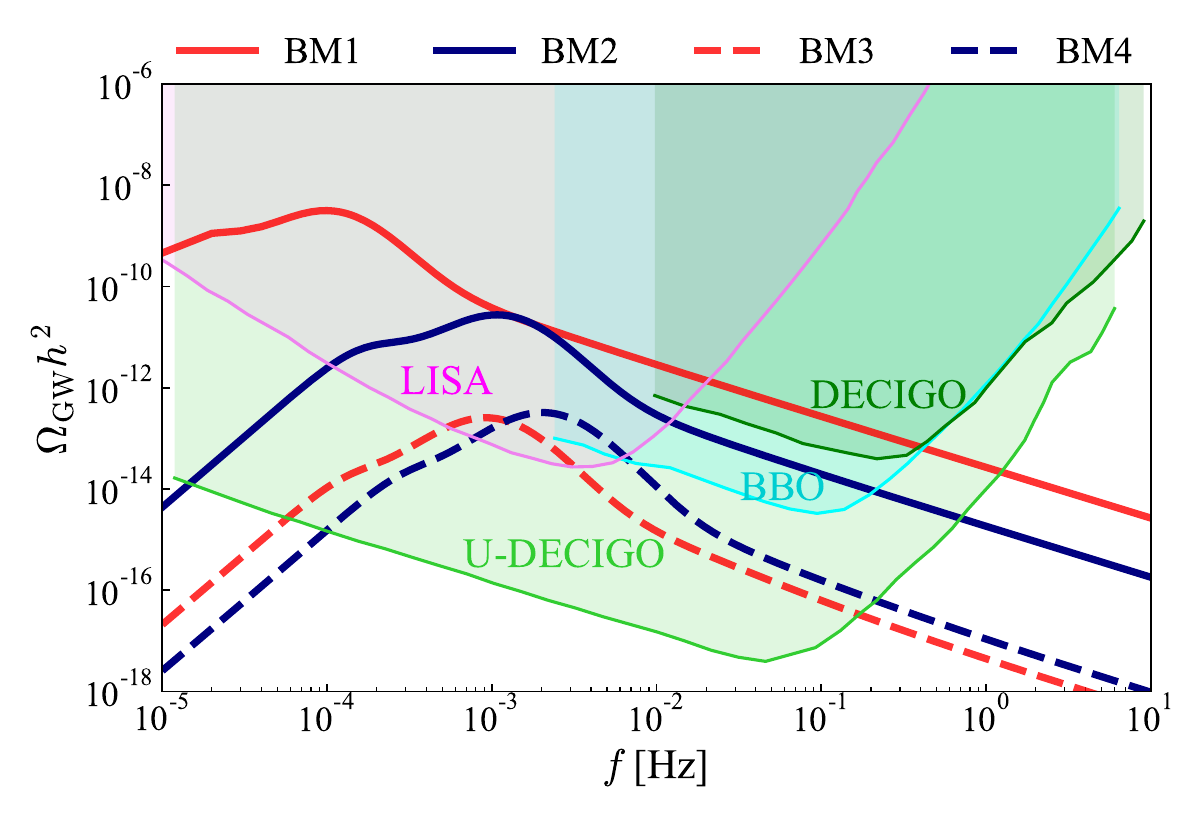}
\caption{GW energy density fraction as a function of the frequency for BM1 (solid red), BM2 (solid blue), BM3 (dotted red), and BM4 (dotted blue). The colored regions are observable areas by future space-based interferometers.}
\label{fig:GW}
 \end{center}
\end{figure}

To see the GW spectrum more specifically, we also show the GW density fraction as a function of the frequency $f$ for all the benchmark parameter sets in Fig.~\ref{fig:GW}.
The spectra for BM1 and BM2 (BM3 and BM4) are indicated by red and blue solid (dotted) lines, respectively. 
For BM3 and BM4, only the second PTs generate the observable spectra in Fig~\ref{fig:GW}.
For BM3 the first-order PT occurs twice; however, the GW spectrum from the first PT is too small to be observed.
The values of $(\alpha, \tilde\beta)$ are $(3.41, 22.4)$, $(2.24, 226)$, $(7.07\times10^{-2}, 55.7)$, and $(0.139, 160)$ for BM1, BM2, BM3, and BM4, respectively.
Because of the low nucleation temperatures for BM1 and BM2, the PT strengths $\alpha$ are sizable, and the peak GW energy density fractions become large, as can be seen in the figure.
We can also see that LISA and U-DECIGO can detect all the GW spectra for benchmark parameter sets even though the suppression factor $\Upsilon$ is involved in the sound wave contribution.
Therefore, we could observe a GW spectrum as a remnant of a PT which influences the relic density of the dark matter as shown in Sec.~\ref{sec:4}.

Another comment is that because of the suppression factor for the sound wave contribution $\Omega_\mathrm{sw}h^2$, 
the contributions from bubble wall collisions $\Omega_\varphi h^2$ become dominant except for the regions around the peaks of the spectra.
Hence, we could see both contributions, $\Omega_\varphi h^2$ and $\Omega_\mathrm{sw} h^2$, in future observations.

Note that if we consider the masses of the inert scalars heavier than $\mathcal{O}(1)$ TeV, the PTs tend to be one-step and become weakly first-order or continuous (second-order or cross-over) 
because the inert scalar masses approach the decoupling limit. 
In this case, the mass shifts for the inert scalars caused by the PTs become small, and the PTs would not affect the dark matter production. 
In fact, a large mass shift for the inert scalars is required to have an impact on the dark matter production, as will be seen in Sec.~\ref{sec:4.1}.

\section{The dark matter relic abundance}
\label{sec:4}
In this section, we examine the impact of PTs on the calculation of the lightest right-handed neutrino dark matter $N_1$ in the scotogenic model. 
Hereafter, the dark matter $N_1$ is simply denoted by $N$. 
At zero temperature, the inert scalar particles $\eta^\pm$, $H$, and $A$ should be heavier than the dark matter $N$, otherwise $N$ becomes unstable and decays into the inert scalar particles. 
The mass ordering may change at finite temperature because of thermal effects. 

\subsection{The case of one-step phase transitions}
\label{sec:4.1}
The benchmark parameter sets BM1 and BM2 lead the one-step PTs. 
In general, we must solve the Boltzmann equations coupled between the dark matter $N$ and the inert scalar particles ($\eta^\pm$, $H$ and $A$) to follow the evolution of the number density of dark matter. 
However we can simplify the Boltzmann equations because some of the inert scalar particles should be almost degenerate to avoid the constraints of the EW precision tests~\cite{Haber:2010bw}. 
Therefore we write down the Boltzmann equations using the notation $n_\eta$ for the number density of the lightest real scalar particle, and $g_{\eta}$ for the degrees of freedom. 
For example, if it is in the thermal bath, the number density is simply given by $n_{\eta}^\mathrm{eq}=K_2(m_{\eta}/T)m_{\eta}^2T/(2\pi^2)$ assuming the Maxwell-Boltzmann distribution. 
Besides, $g_\eta=4$ if all the inert scalar particles $\eta^{\pm}$, $H$ and $A$ are degenerate, whereas $g_\eta=3$ if only $\eta^{\pm}$ and $A$ are degenerate and lightest, 
or $g_\eta=1$ if $H$ is the lightest particle. 
With this simplification, we can write down the Boltzmann equations for the dark matter $N$ coupled with the inert scalar particles as follows:
\begin{align}
\frac{dn_N}{dt}+3Hn_N&=
g_\eta\langle\Gamma_\eta\rangle\left(n_\eta-n_\eta^\mathrm{eq}\frac{n_N}{n_N^\mathrm{eq}}\right)
-g_\eta\langle\Gamma_N\rangle\left(n_N-n_N^\mathrm{eq}\frac{n_{\eta}}{n_{\eta}^\mathrm{eq}}\right),
\label{eq:boltz1}\\
%%%
\frac{dn_\eta}{dt}+3Hn_\eta&=-g_{\eta}\langle\sigma_{\eta\eta}v_\mathrm{rel}\rangle\left(n_\eta^2-{n_\eta^\mathrm{eq}}^2\right)
-\langle\Gamma_\eta\rangle\left(n_\eta-n_\eta^\mathrm{eq}\frac{n_N}{n_N^\mathrm{eq}}\right)
+\langle\Gamma_N\rangle\left(n_N-n_N^\mathrm{eq}\frac{n_{\eta}}{n_{\eta}^\mathrm{eq}}\right),
\label{eq:boltz2}
\end{align}
where $\langle\sigma_{\eta\eta}v_\mathrm{rel}\rangle$ is the thermally averaged annihilation cross section of the lightest inert scalar particle, 
$\langle\Gamma_\eta\rangle$ and $\langle\Gamma_N\rangle$  are the decay widths of $\eta$ and $N$ with thermal effects, which are given by
\begin{align}
\langle\Gamma_\eta\rangle&=\frac{|\mathbb{Y}|^2}{16\pi}m_{\eta}\left(1-\frac{m_N^2}{m_\eta^2}\right)^{2}\frac{K_1(m_\eta/T)}{K_2(m_\eta/T)}~\theta(m_{\eta}-m_N),
\label{eq:eta_decay}\\
\langle\Gamma_N\rangle&=\frac{|\mathbb{Y}|^2}{32\pi}m_{N}\left(1-\frac{m_\eta^2}{m_N^2}\right)^{2}\frac{K_1(m_N/T)}{K_2(m_N/T)}~\theta(m_N-m_{\eta}).
\label{eq:N_decay}
\end{align}
In the above equations, $|\mathbb{Y}|\equiv\sqrt{|y_{1e}|^2+|y_{1\mu}|^2+|y_{1\tau}|^2}$, 
$K_n(x)$ is the second kind of the modified Bessel function with order $n$, $\theta(x)$ is the step function, and the light lepton masses are ignored. 

The annihilation cross section $\langle\sigma_{\eta\eta}v_\mathrm{rel}\rangle$ can numerically be evaluated by using micrOMEGAs~\cite{Belanger:2018ccd}. 
The temperature dependence of the cross section can safely be ignored because the inert scalar particles decouple from the thermal bath when they are non-relativistic ($m_{\eta}/T\sim25$). 
Therefore we use the fixed value of the cross section at zero temperature in this work, which is typically much larger than the canonical value $3\times10^{-26}~\mathrm{cm^3/s}$ 
for thermal relic abundance because of the certain size of gauge interactions.

When the above Boltzmann equations (\ref{eq:boltz1}) and (\ref{eq:boltz2}) are derived, 
the dark matter $N$ is implicitly assumed to be in kinetic equilibrium with the thermal bath particles, 
namely $f_N\propto f_N^\mathrm{eq}$ is assumed. 
This is valid if the elastic scattering processes between $N$ and the thermal bath particles are fast enough. 
This is true in the current model because the elastic scattering process $NL_\alpha\to\eta^*\to NL_\alpha$ is enhanced by the $\eta$ resonance even if the neutrino Yukawa coupling is small.

\vspace{0.5cm}
The set of the Boltzmann equations in Eqs.~(\ref{eq:boltz1}) and (\ref{eq:boltz2}) can be rewritten as
\begin{align}
\frac{dY_N}{dx}=&~\frac{g_{\eta}\langle\Gamma_{\eta}\rangle}{x\tilde{H}}\left[Y_{\eta}-\frac{Y_{\eta}^\mathrm{eq}Y_N}{Y_N^\mathrm{eq}}\right]
-\frac{g_{\eta}\langle\Gamma_N\rangle}{x\tilde{H}}\left[Y_N-\frac{Y_N^\mathrm{eq}Y_\eta}{Y_\eta^\mathrm{eq}}\right],
\label{eq:boltz1-2}\\
\frac{dY_\eta}{dx}=&-\frac{g_{\eta}\langle\sigma_{\eta\eta}v_\mathrm{rel}\rangle s}{x\tilde{H}}\left[Y_\eta^2-{Y_\eta^\mathrm{eq}}^2\right]
-\frac{g_{\eta}\langle\Gamma_{\eta}\rangle}{x\tilde{H}}\left[Y_{\eta}-\frac{Y_{\eta}^\mathrm{eq}Y_N}{Y_N^\mathrm{eq}}\right]
+\frac{g_{\eta}\langle\Gamma_N\rangle}{x\tilde{H}}\left[Y_N-\frac{Y_N^\mathrm{eq}Y_\eta}{Y_\eta^\mathrm{eq}}\right],
\label{eq:boltz2-2}
\end{align}
where $x\equiv m_N/T$ is the dimensionless parameter given by the temperature of the universe $T$, which is related to the time, 
and $Y_{N,\eta}\equiv n_{N,\eta}/s$ is the yield written by the entropy density $s=2\pi^2g_{*s}T^3/45$. 
The modified Hubble rate $\tilde{H}$ is given by
\begin{align}
\tilde{H}\equiv H\left(1-\frac{x}{3g_{*s}}\frac{dg_{*s}}{dx}\right)^{-1}, 
\end{align}
with the effective relativistic degrees of freedom $g_{*s}$ for the entropy density. 
In the above equations, we include thermal effects in the particle masses, as discussed in the previous section. 
The dark matter relic abundance should satisfy the value $\Omega_N h^2=0.12$ observed by PLANCK~\cite{Planck:2018vyg}, and this is translated into the yield at $m_N/T\to\infty$ as 
\begin{align}
Y_N^\infty=4.4\times10^{-11}\left(\frac{\mathrm{GeV}}{m_N}\right). 
\end{align}

\begin{figure}[t]
 \begin{center}
  \includegraphics[width=6.9cm]{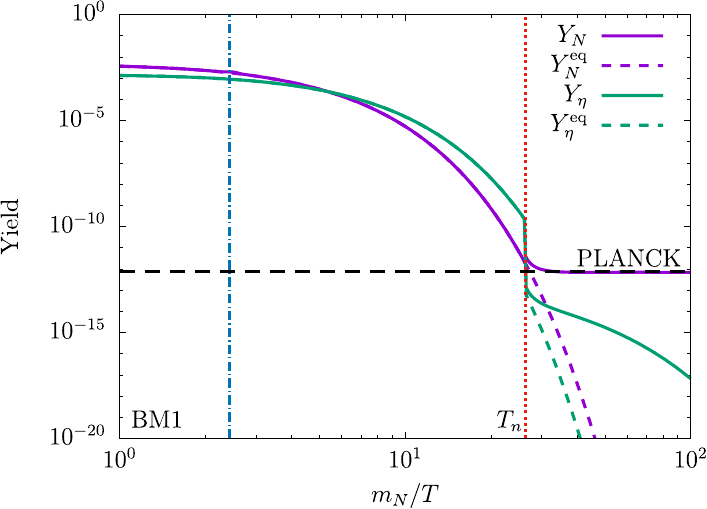}
  \includegraphics[width=6.9cm]{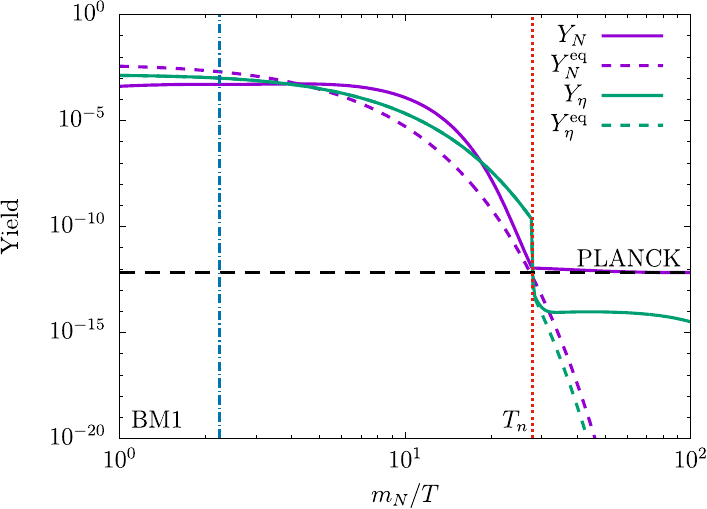}\\
  \hspace{0.19cm}
  \includegraphics[width=6.6cm]{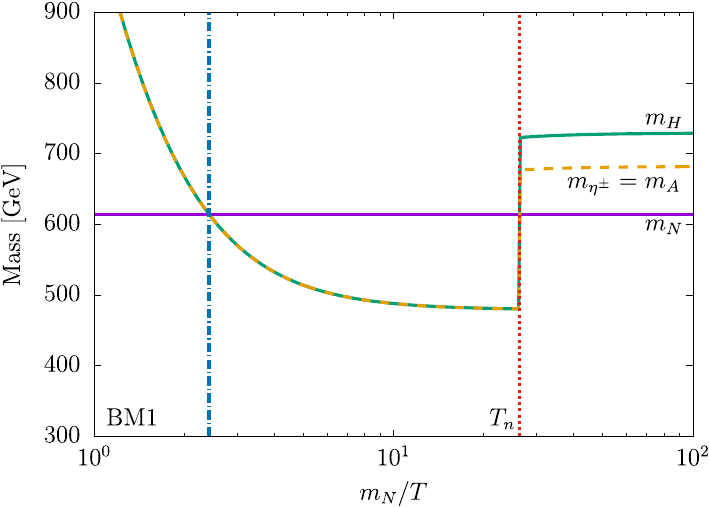}
  \hspace{0.16cm}
  \includegraphics[width=6.6cm]{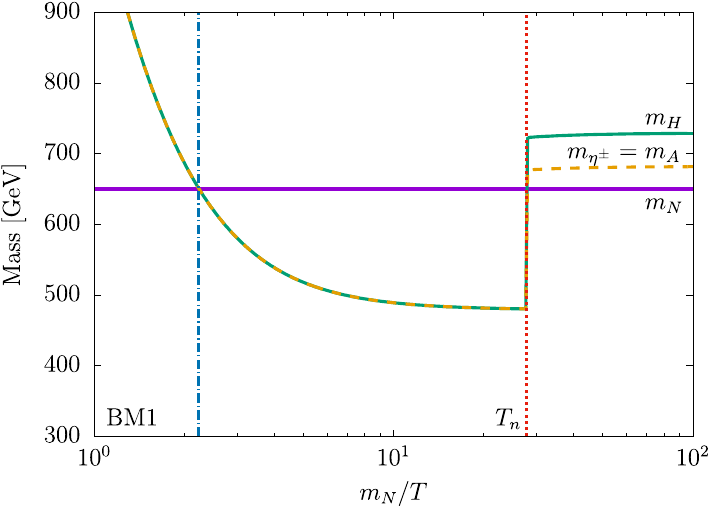}\\
  \includegraphics[width=6.9cm]{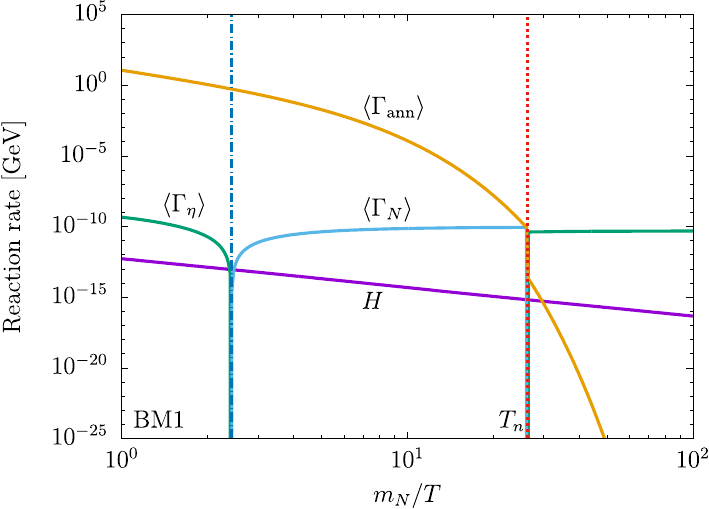}
  \includegraphics[width=6.9cm]{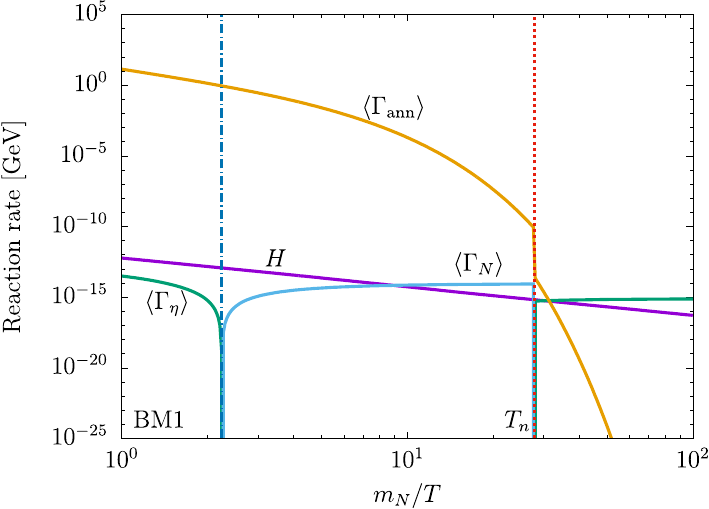}
\caption{Example solutions of the coupled Boltzmann equations (top), the scalar masses with thermal effects (middle), and the reaction rates (bottom) for the one-step PT (BM1) where 
the dark matter mass and the Yukawa coupling are fixed at $(m_N,~|\mathbb{Y}|)=(614~\mathrm{GeV},~1.00\times10^{-5})$ for the left panels and $(650~\mathrm{GeV},~8.20\times10^{-8})$ for 
the right panels, respectively. }
\label{fig:examples}
 \end{center}
\end{figure}

\begin{figure}[t]
 \begin{center}
  \includegraphics[width=6.9cm]{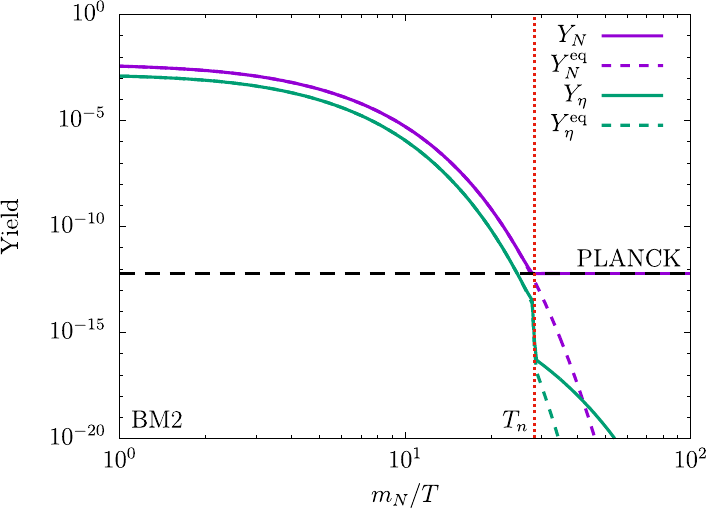}
  \includegraphics[width=6.9cm]{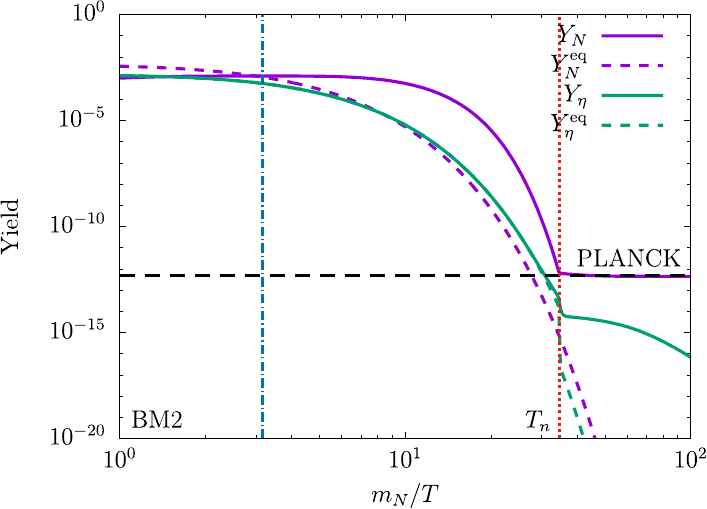}\\
  \hspace{0.05cm}
  \includegraphics[width=6.8cm]{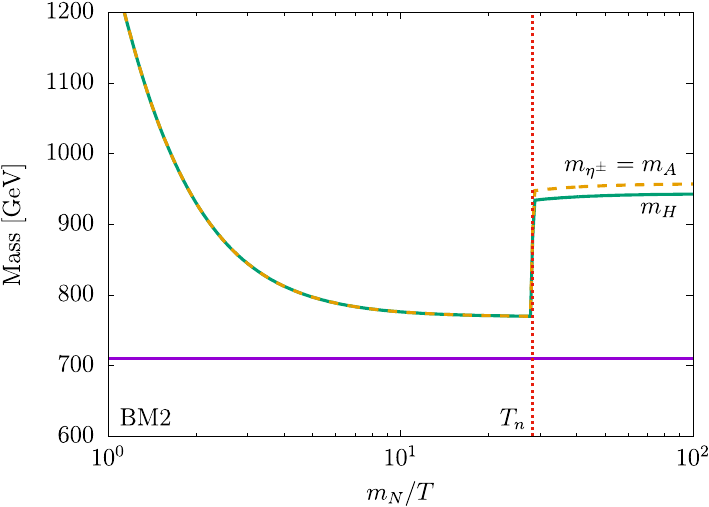}
  \hspace{-0.05cm}
  \includegraphics[width=6.8cm]{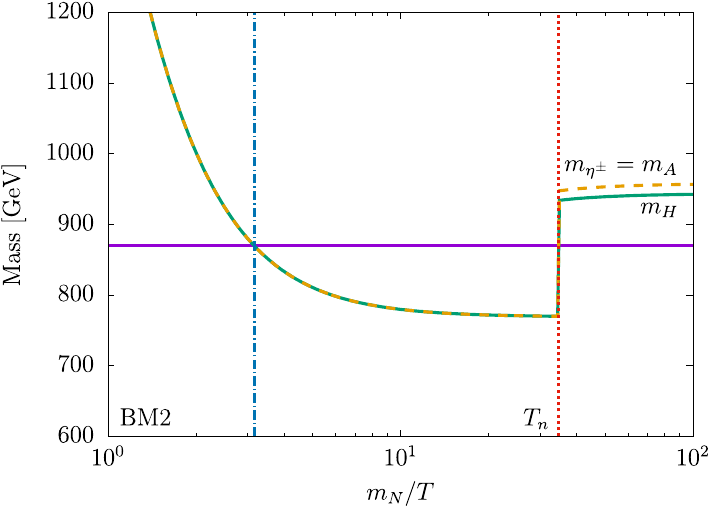}\\
  \includegraphics[width=6.9cm]{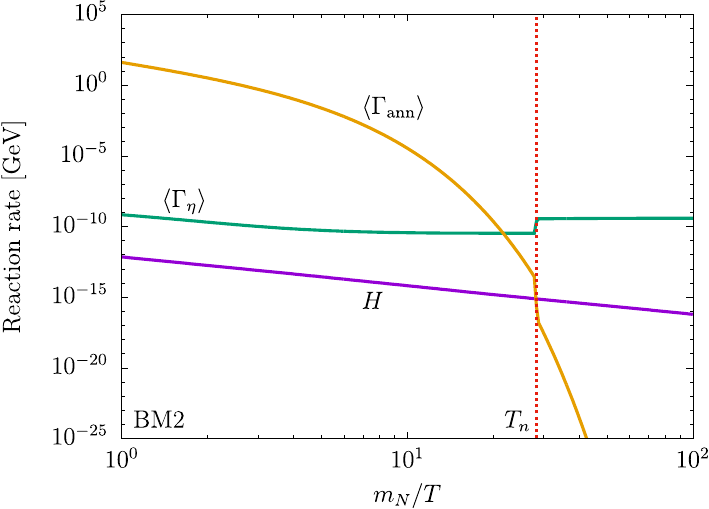}
  \includegraphics[width=6.9cm]{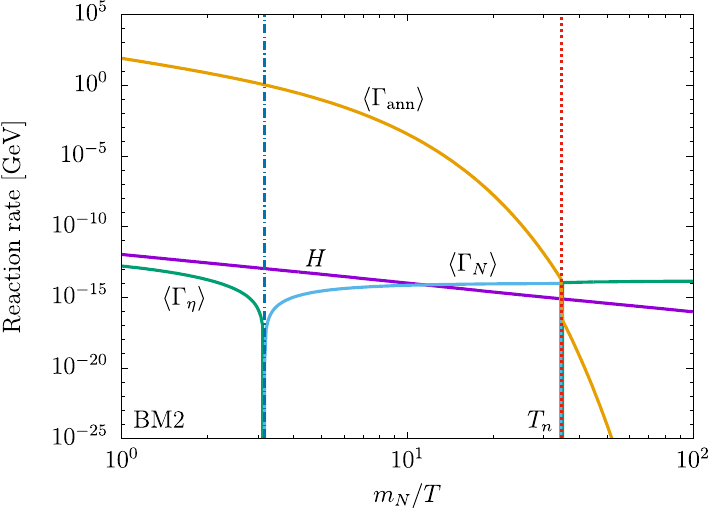}
\caption{Same plots with Fig.~\ref{fig:examples} for BM2 where the dark matter mass and the Yukawa coupling are fixed at $(m_N,~|\mathbb{Y}|)=(710~\mathrm{GeV},~1.00\times10^{-5})$ for the left panels, 
and $(870~\mathrm{GeV},~1.54\times10^{-7})$ for the right panels, respectively. }
\label{fig:examples2}
 \end{center}
\end{figure}

Two example solutions for Eqs.~(\ref{eq:boltz1-2}) and (\ref{eq:boltz2-2}) are shown in Fig.~\ref{fig:examples} for BM1 setting the initial number densities $Y_{N,\eta}^\mathrm{init}=0$ 
where the dark matter mass and the Yukawa coupling are fixed to be $(m_N,|\mathbb{Y}|)=(614~\mathrm{GeV},~1.00\times10^{-5})$ for the left panels and $(650~\mathrm{GeV},~8.20\times10^{-8})$ 
for the right panels.
The top, middle, and bottom panels represent the evolution of the number densities, the masses with thermal effects, and the reaction rates for the decay 
and the $\eta$ annihilation rate defined by $\langle\Gamma_\mathrm{ann}\rangle\equiv\langle\sigma_{\eta\eta}{v}_\mathrm{rel}\rangle n_\eta^\mathrm{eq}$ with the Hubble rate, respectively. 
In the top panels, the horizontal black dashed line represents the dark matter relic abundance observed by PLANCK~\cite{Planck:2018vyg}. 
In all the panels, the vertical dark-blue dash-dotted line denotes the temperature that the mass flip between $N$ and $\eta$ occurs, while the red dotted line denotes the nucleation temperature $T_n$. 

In the left panels, one can see that the decay and annihilation processes of $\eta$ are initially thermalized ($\langle\Gamma_\eta\rangle,\langle\Gamma_\mathrm{ann}\rangle>H$), 
thus the yield $Y_{N,\eta}$ follows the equilibrium values $Y_{N,\eta}=Y_{N,\eta}^\mathrm{eq}$. 
From the left middle plot, one can find that the scalar masses become smaller as the temperature decreases. 
At $x\approx2.4~(T\approx254~\mathrm{GeV})$, the masses between $N$ and $\eta$ turn over. 
Thus after this point, the $N$ decay becomes active instead of the $\eta$ decay. 
The masses between $N$ and $\eta$ turn over again at $x\approx26~(T\approx23~\mathrm{GeV})$ where the first-order EWPT occurs ($\phi_1\neq0$). 
This is because the $\eta$ masses obtain a large correction from $\phi_1$ (see the left middle plot). 
At this point, the reaction rate for the $\eta$ annihilation drastically decreases and becomes ineffective shortly afterwards. 
In addition, because the $\eta$ decay becomes effective again, a large number of $\eta$ is converted into $N$ via the $\eta$ decay. 
Simultaneously, the inverse decay also competes with the $\eta$ decay, and the dark matter number density settles in abundance observed by PLANCK at the end~\cite{Planck:2018vyg}.

In the right panels, $\eta$ is initially in a thermal bath, whereas $N$ is not. 
After the mass flip at $x\approx2.2~(T\approx290~\mathrm{GeV})$, the $N$ number density starts to slowly decrease via the $N$ decay. 
The $N$ yield is instantaneously fixed when the EWPT occurs at $x\approx28~(T\approx23~\mathrm{GeV})$ 
because the $\eta$ mass suddenly obtains a large correction as same as the left panels. 
As a result, the masses between $N$ and $\eta$ turn over. 
Unlike the left panels, the $N$ number density does not change drastically at this point because the decay width $\langle\Gamma_\eta\rangle$ is much smaller than in the previous example 
due to the smaller Yukawa coupling.

The other two example solutions for BM2 are shown in Fig.~\ref{fig:examples2} where the dark matter mass and the Yukawa coupling are fixed at 
$(m_N,|\mathbb{Y}|)=(710~\mathrm{GeV},1.00\times10^{-5})$ for the left panels and $(870~\mathrm{GeV},1.54\times10^{-7})$ for the right panels. 
As same as the previous examples in Fig.~\ref{fig:examples}, the dark matter relic abundance is instantaneously determined at the nucleation temperature $T_n\approx25~\mathrm{GeV}$. 
In the left panels, the mass flipping between $N$ and $\eta$ does not occur, and the dark matter $N$ is the lightest all the time. 
Even in this case, the $N$ number density is reduced by the inverse decay $NL_\alpha\to\eta$, and the observed relic abundance can be reproduced.

\begin{figure}[t]
 \begin{center}
  \includegraphics[width=7.5cm]{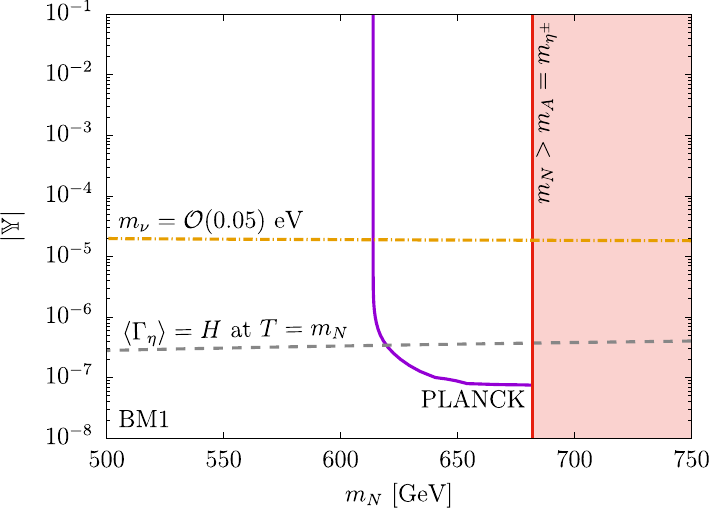}
  \includegraphics[width=7.5cm]{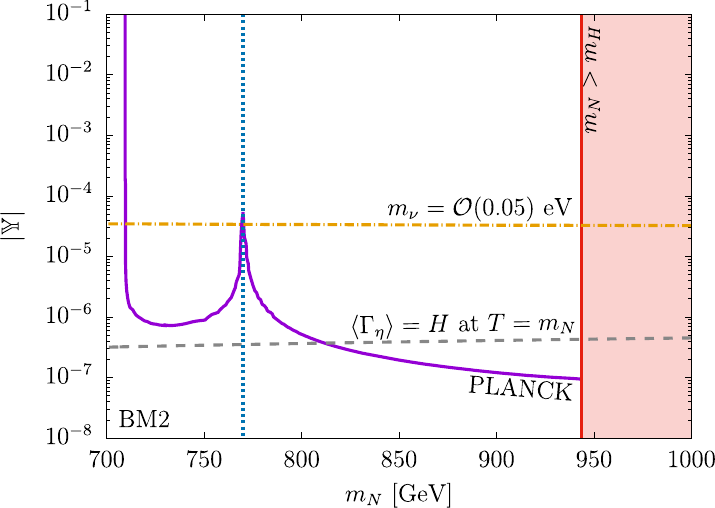}
\caption{Parameter space in the $(m_N,|\mathbb{Y}|)$ plane for BM1 (left) and BM2 (right). 
  The purple lines can reproduce the observed relic abundance. 
In the red regions on the right-handed side, the dark matter candidate $N$ becomes heavier than the inert scalars. 
The orange dashed-dotted line corresponds to the order of the neutrino Yukawa coupling which can predict the correct small neutrino mass scale $\mathcal{O}(0.05)~\mathrm{eV}$. 
  The vertical blue dotted line around $m_N\sim770~\mathrm{GeV}$ in the right panel corresponds to the threshold $m_N= m_{\eta}$ for $T\gtrsim T_n$. 
The gray dashed lines the lines of $\langle\Gamma_{\eta}\rangle=H$ at $T=m_N$. 
}
\label{fig:contour}
 \end{center}
\end{figure}

\vspace{0.5cm}
Fig.~\ref{fig:contour} shows the parameter space in the ($m_N$, $|\mathbb{Y}|$) plane where the purple lines can reproduce the observed relic abundance for the benchmark BM1 (left) and BM2 (right). 
The red regions on the right-hand side are not appropriate for our work because the lightest right-handed neutrino $N$ becomes heavier than the lightest inert scalar particle at zero temperature. 
The orange dash-dotted line corresponds to the size of the neutrino Yukawa coupling such that the predicted neutrino mass scale given by Eq.~(\ref{eq:nu-mass}) is $\mathcal{O}(0.05)~\mathrm{eV}$, 
which is inferred by the neutrino oscillation experiments~\cite{deSalas:2020pgw}. 
The gray dashed line represents the contour that the decay rate $\langle\Gamma_{\eta}\rangle$ can be comparable to the Hubble rate ($\langle\Gamma_\eta\rangle\sim H$) at $T=m_N$. 
This gives a criterion that the decays occur fast enough or not compared to the universe expansion. 
The vertical dark-blue dotted line in the right panel corresponds to the threshold $m_N=m_{\eta}$ for $T\gtrsim T_n$. 
From the figure, one can find that the dark matter relic abundance does not depend on the size of the Yukawa coupling $|\mathbb{Y}|$ 
when the dark matter is $m_N\sim 614~\mathrm{GeV}$ (left) and $m_N\sim 710~\mathrm{GeV}$ (right). 
In addition, the dependence on the dark matter mass becomes small when the mass is close to the mass threshold (red region), 
and thus the relic abundance is mainly determined by the Yukawa coupling $|\mathbb{Y}|$. 
These behaviours coincide with the previous work~\cite{Baker:2018vos}. 
%%%
From these numerical calculations, we find that the observed dark matter relic abundance is almost fixed at the temperature that the EWPT occurs ($T=T_n$) 
as in the examples we have seen in Figs.~\ref{fig:examples} and \ref{fig:examples2}.

\vspace{0.5cm}
We comment on the charged lepton flavor violating process $\mu\to e\gamma$ here~\cite{Toma:2013zsa}. 
When the Yukawa couplings are smaller than $\mathcal{O}(0.05)$, the current experimental bound can be avoided~\cite{ParticleDataGroup:2020ssz}, and 
the predicted branching fraction for $\mu\to e\gamma$ can be testable by future experiments. 
In the previous works~\cite{Kubo:2006yx, Suematsu:2009ww, Schmidt:2012yg}, it has been found that 
the thermal production of right-handed neutrino dark matter tends to conflict with the strong constraint of the charged flavor violating processes. 
However, the first-order PTs can largely affect the production of dark matter in some specific parameter sets, as have been shown in this work.
This makes it possible for the model to be consistent with the constraints and to test with future experiments. 
In addition, if the Yukawa couplings are complex, the electric dipole moments of the charged leptons are predicted~\cite{Abada:2018zra, Fujiwara:2020unw, Fujiwara:2021vam}, which may be related to 
a generation of the baryon asymmetry in the universe. 

Another comment is that if the inert scalar masses are heavier than TeV scale, the mass shift via the EWPT would be small as $\mathcal{O}(10)\%$ 
as can be seen from Eqs.~(\ref{eq:etapm})--(\ref{eq:A}). 
This is because the VEV $v$ is the EW scale, and the quadratic term in the potential $\mu_\eta$ should be $\mu_\eta\gtrsim\mathcal{O}(1)~\mathrm{TeV}$.  
In this case, the PT effects on the calculation of dark matter relic abundance would be restrictive.

\subsection{The case of two-step phase transitions}
For two-step PTs, the inert scalar doublet $\eta$ can temporarily have the VEV $(\phi_2\neq0)$ during $T_{n2}<T<T_{n1}$ 
where $T_{n1}$ and $T_{n2}$ are the nucleation temperatures that the first and second PTs occur, respectively. 
As a result, in the interval $T_{n2}<T<T_{n1}$ the left-handed and right-handed neutrinos mix with each other through the neutrino Yukawa coupling $y_{i\alpha}$. 
Thus additional decay channels $W^{\pm}\to N\ell_\alpha^{\pm}$ and $Z\to N\nu_{\alpha}$ can occur, which may contribute to the evolution of the dark matter number density.

In this section, we focus on the case that the dark matter $N$ is much lighter than the inert scalar particles ($m_N\ll m_{\eta^\pm}, m_H, m_A$), and is never thermalized with the other particles to investigate the impact of the PTs on the calculation of dark matter relic abundance.\footnote{One can also consider the other possibilities such as $m_N\gtrsim m_{\eta^\pm}, m_H, m_A$ and dark matter thermalized cases as same as the previous section. However the impact of the PTs on the dark matter relic abundance is restrictive in these cases.} 
Because the inert scalar particles $\eta^{\pm}, H$, and $A$ are in thermal equilibrium during the evolution of dark matter number density ($n_\eta=n_\eta^\mathrm{eq}$), the Boltzmann equation for the dark matter is simply given by
\begin{align}
\frac{dn_N}{dt}+3Hn_N=
g_\eta\langle\Gamma_\eta\rangle n_\eta^\mathrm{eq}
+\Big(\langle\Gamma_{W_T}\rangle n_{W_T}^\mathrm{eq} + \langle\Gamma_{W_L}\rangle n_{W_L}^\mathrm{eq}\Big)
+\Big(\langle\Gamma_{Z_T}\rangle n_{Z_T}^\mathrm{eq} + \langle\Gamma_{Z_L}\rangle n_{Z_L}^\mathrm{eq}\Big),
\label{eq:boltz3}
\end{align}
where the first term on the right-hand side is the contribution from the decay channel $\eta\to NL_\alpha$ given by Eq.~(\ref{eq:eta_decay}), whereas the other terms are the new contributions corresponding to the gauge boson decays coming from the temporary VEV $\phi_2$. 
Because we include the finite temperature effect in the masses, the gauge covariance is lost, and thus the transverse and longitudinal modes have different masses. 
Therefore we must calculate the decay widths for the transverse and longitudinal modes separately. 
This is shown in Eq.~(\ref{eq:boltz3}) with the subscripts $T$ and $L$ denoting the transverse and longitudinal modes. 
Each decay width is calculated as
\begin{align}
 \langle\Gamma_{W_T}\rangle\approx&~\frac{2}{3}\frac{g^2m_{W_T}}{32\pi}|\mathbb{Y}|^2\frac{\phi_2^2}{m_N^2}
\frac{K_1\left(m_{W_T}/T\right)}{K_2\left(m_{W_T}/T\right)}\theta(T_{n1}-T)\theta(T-T_{n2}),
\label{eq:WT}\\
 \langle\Gamma_{W_L}\rangle\approx&~\frac{1}{3}\frac{g^2m_{W_L}}{16\pi}|\mathbb{Y}|^2\frac{\phi_2^2}{m_N^2}
\frac{K_1\left(m_{W_L}/T\right)}{K_2\left(m_{W_L}/T\right)}\theta(T_{n1}-T)\theta(T-T_{n2}),
\label{eq:WL}\\
 \langle\Gamma_{Z_T}\rangle\approx&~\frac{2}{3}\frac{g^2m_{Z_T}}{32\pi\cos^2\theta_W}|\mathbb{Y}|^2\frac{\phi_2^2}{m_N^2}
\frac{K_1\left(m_{Z_T}/T\right)}{K_2\left(m_{Z_T}/T\right)}\theta(T_{n1}-T)\theta(T-T_{n2}),
\label{eq:ZT}\\
 \langle\Gamma_{Z_L}\rangle\approx&~\frac{1}{3}\frac{g^2m_{Z_L}}{16\pi\cos^2\theta_W}|\mathbb{Y}|^2\frac{\phi_2^2}{m_N^2}
\frac{K_1\left(m_{Z_L}/T\right)}{K_2\left(m_{Z_L}/T\right)}\theta(T_{n1}-T)\theta(T-T_{n2}),
\label{eq:ZL}
\end{align}
where the gauge boson masses are assumed to be much heavier than leptons and dark matter $N$. 
Note that the factor $\phi_2^2/m_N^2$ appears in the above equations, which comes from the neutrino mixing matrix elements.
This factor enhances the gauge boson decay widths if the dark matter mass is lighter than the VEV $\phi_2$. 
One can confirm that the second and third terms in the right-hand side of Eq.~(\ref{eq:boltz3}) are simply reduced to 
$\langle\Gamma_W\rangle n_W^\mathrm{eq}$ and $\langle\Gamma_Z\rangle n_Z^\mathrm{eq}$, respectively, if the finite temperature effect is not included in the gauge boson masses. 
This is because the transverse and longitudinal modes have the same masses in this limit.

\begin{figure}[t]
 \begin{center}
  \includegraphics[width=7.5cm]{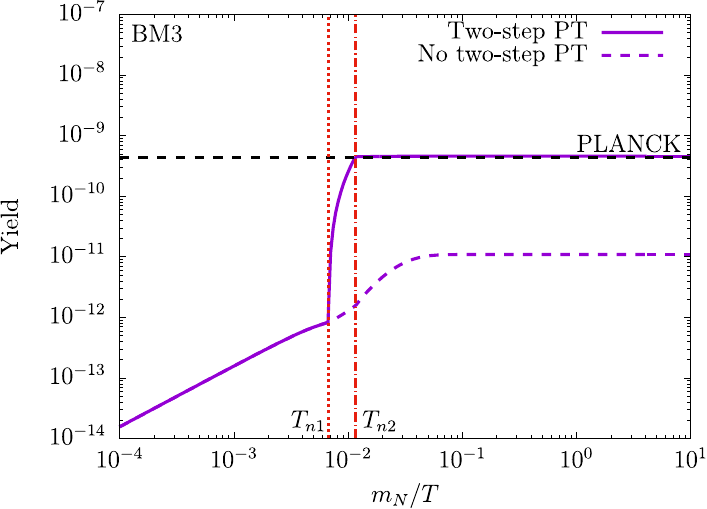}
  \includegraphics[width=7.5cm]{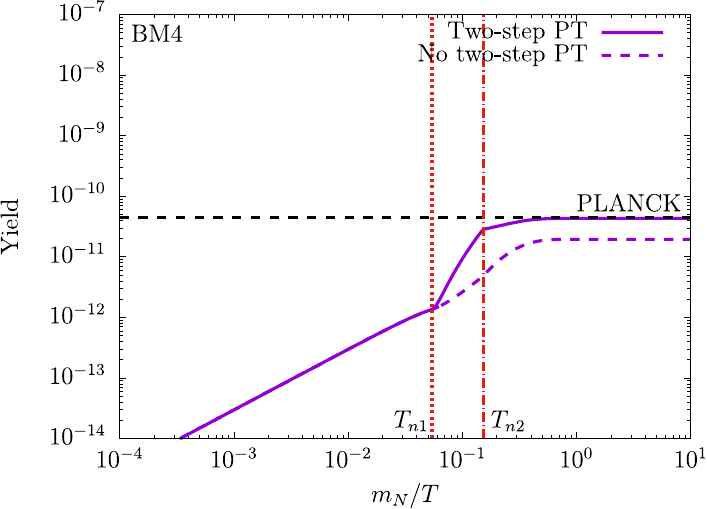}\\
  \includegraphics[width=7.5cm]{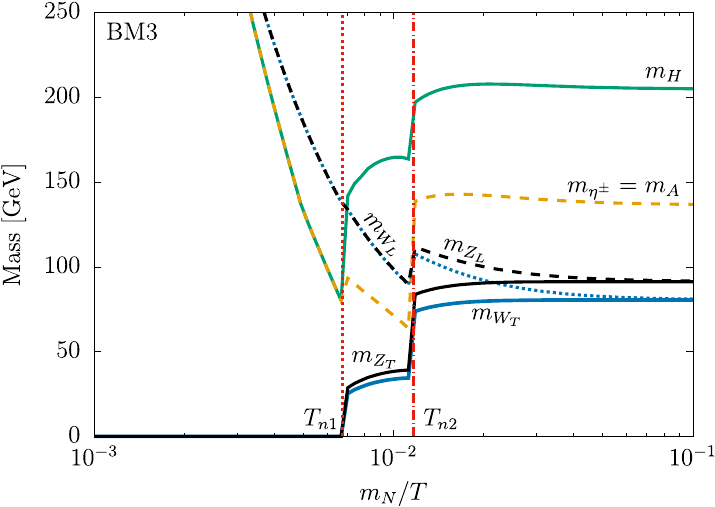}
  \includegraphics[width=7.5cm]{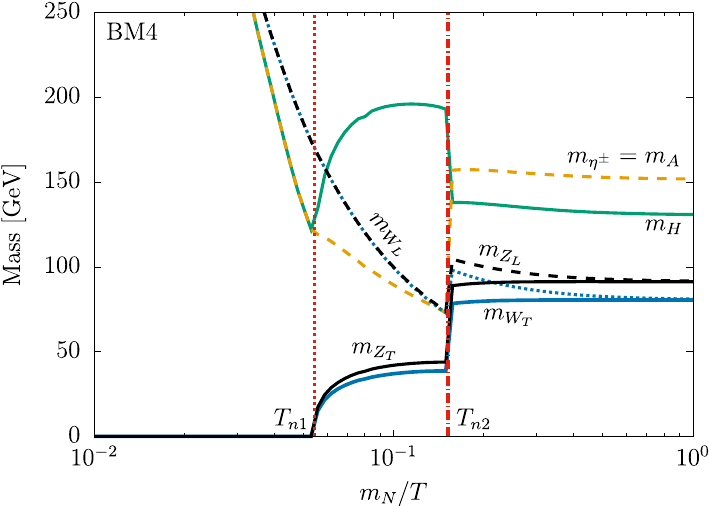}
\caption{Example solutions for the Boltzmann equation with the two-step PTs (top) and the masses of the inert scalars and gauge bosons with thermal effects (bottom). 
The left and right panels correspond to BM3 and BM4, respectively. 
The dark matter mass and the Yukawa coupling are fixed at $(m_N,~|\mathbb{Y}|)=(1~\mathrm{GeV},~2.1\times10^{-12})$ for the left panels and $(10~\mathrm{GeV},~2.8\times10^{-12})$ 
for the right panels.}
\label{fig:examples_2step}
 \end{center}
\end{figure}

Two example solutions for the Boltzmann equation, and masses of the inert scalar and gauge bosons are shown in Fig.~\ref{fig:examples_2step} for BM3 (left) and BM4 (right). 
The dark matter mass and the neutrino Yukawa coupling are fixed at $m_N=1~\mathrm{GeV}$ and $|\mathbb{Y}|=2.1\times10^{-12}$ in the left panels, and $m_N=10~\mathrm{GeV}$ 
and $|\mathbb{Y}|=2.8\times10^{-12}$ in the right panels. 
In the top panels, the solid purple lines represent the solutions for the benchmark parameter sets, whereas the dashed purple lines are the fictional solutions without the two-step PTs, 
namely no contribution from the gauge boson decays in Eq.~(\ref{eq:boltz3}), which are shown for comparison. 
The horizontal black dotted line in each plot is the dark matter relic abundance observed by PLANCK~\cite{Planck:2018vyg}. 
The vertical red dotted and dash-dotted lines represent the nucleation temperatures $T_{n1}$ and $T_{n2}$. 
We can find the large enhancement of dark matter number density during the period $T_{n2}<T<T_{n1}$. 
This is because the production rate for the dark matter is enhanced by the factor $\phi_2^2/m_N^2$ as can be seen in Eq.~(\ref{eq:WT}) -- (\ref{eq:ZL}). 
Note again that we focused on the parameter space different from the one-step PT cases for BM1 and BM2 to investigate the impact on the calculation of dark matter relic abundance.

\begin{figure}[t]
 \begin{center}
  \includegraphics[width=7.5cm]{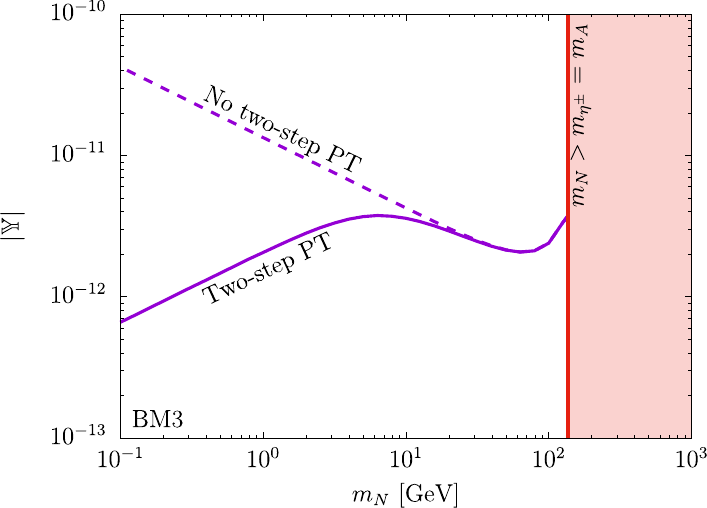}
  \includegraphics[width=7.5cm]{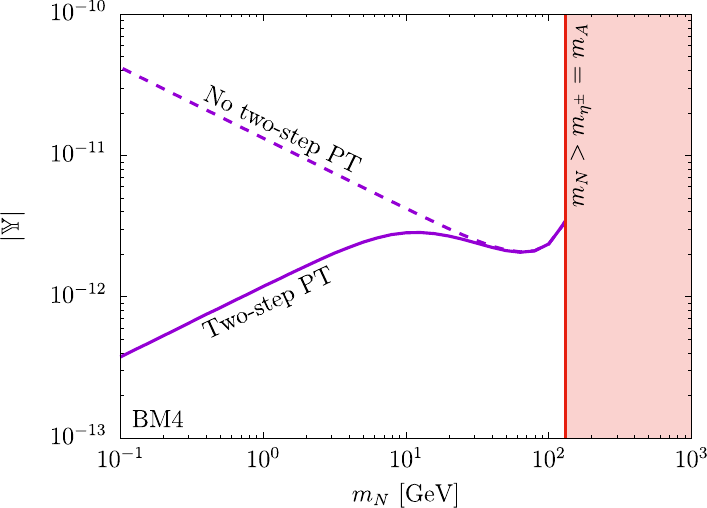}
\caption{Contours reproducing the observed relic abundance for BM3 and BM4.}
\label{fig:contour_2step}
 \end{center}
\end{figure}

Fig.~\ref{fig:contour_2step} shows the parameter space which can reproduce the correct relic abundance in the ($m_N, |\mathbb{Y}|$) plane for BM3 (left) and BM4 (right).
In the red region on the right-hand side, the dark matter candidate $N$ becomes heavier than the inert scalar particles. 
We can find that when $m_N\gtrsim\mathcal{O}(100)~\mathrm{GeV}$ the contribution from the gauge boson decays due to the two-step PTs is small and has no impact on the dark matter relic abundance, 
whereas it gives a large impact when $m_N\lesssim\mathcal{O}(10)~\mathrm{GeV}$. 
This is because the gauge boson decay widths given by Eq.~(\ref{eq:WT}) -- (\ref{eq:ZL}) are proportional to the factor $\phi_2^2/m_N^2$. 

\section{Conclusion}
\label{sec:5}
The nature of the dark matter in the universe is still unknown. 
The production mechanisms for dark matter may be a hint for understanding the universe. 
The freeze-out and freeze-in mechanisms are known as the standard ways to produce dark matter. 
In this work, we investigated the impact of the PTs in the early universe on the production of dark matter in the scotogenic model, 
which is extended by an inert doublet scalar and three right-handed neutrinos (singlet fermions). 
This model is economical for accommodating the small neutrino masses at the one-loop level and involving dark matter candidates simultaneously. 
It is known that the singlet fermionic dark matter candidate in the model tends to be excluded by the relevant constraints if the thermal production via the freeze-out mechanism is assumed. 

At first, we formulated the finite temperature effective potential at the one-loop level with thermal masses. 
Then, we investigated the PTs and performed the parameter search in this model. 
We chose four different benchmark parameter sets inducing the one-step and two-step first-order PTs. 
The first-order PTs generate gravitational waves, which the future space-based interferometers can examine. 

Subsequently, we incorporated the one-step and two-step PTs in the calculations of the dark matter number density in four benchmark parameter sets. 
In the case of one-step PTs, we found that the dark matter number density is instantaneously determined at the nucleation temperature $T_{n}$. 
If the nucleation temperature is low as $T_n \approx 23\sim25~\mathrm{GeV}$, the observed relic abundance can appropriately be reproduced 
with a moderate size of the neutrino Yukawa coupling from $\mathcal{O}(10^{-7})$ to $\mathcal{O}(10^{-1})$. 
Moreover, because of such low nucleation temperatures, the PT becomes very strong first-order and generates the GWs that future experiments can explore.
Therefore we emphasize that the observable GWs and the dark matter production are closely corrected in the one-step PTs.

In the case of two-step PTs, the inert scalar doublet has a temporary VEV only in the interval $T_{n2}<T<T_{n1}$. 
This VEV changes the gauge boson masses and induces the mixing between the left-handed and right-handed neutrinos. 
Thus the new decay channels relevant to the dark matter production appeared through the neutrino mixing. 
These new channels modified the freeze-in mechanism of dark matter production if the dark matter is much lighter than the VEV of the inert scalar. 
Similar to the one-step case, future GW interferometers may clarify such two-step scenarios.

\section*{Acknowledgments}
HS would like to thank Koichi Funakubo for the valuable discussion. 
The numerical computation in this work was carried out at the Yukawa Institute Computer Facility.
This work was supported by JST SPRING, Grand No.~JPMJSP2135 (HS), and JSPS Grant-in-Aid for Scientific Research KAKENHI Grant No. JP20K22349 (TT).

\end{document}